\documentclass{aa}

\usepackage{graphicx}
\usepackage{txfonts}
\usepackage{comment}
\usepackage{dblfloatfix} 

\usepackage{xcolor}
\usepackage{csquotes}
\usepackage{cancel}

\usepackage{booktabs}
\usepackage{multirow}
\usepackage{array}

\begin{document} 

   \title{
   Radio emission in star-forming galaxies: connection to restarted or relic AGN activity}
   \titlerunning{BPT-selected AGN in MaNGA}

   \author{M. Alb\'an \inst{1}
          \and
          D. Wylezalek\inst{1}
          \and P. Kukreti\inst{1}
          \and R. A. Riffel\inst{2}
          \and R. Riffel\inst{3}
          }

   \institute{
   \inst{1}Zentrum f\"ur Astronomie der Universit\"at Heidelberg, Astronomisches Rechen-Institut, M\"onchhofstr, 12-14 69120 Heidelberg, Germany\\
      \inst{2}Departamento de Física, Centro de Ciências Naturais e Exatas, Universidade Federal de Santa Maria, 97105-900, Santa Maria, RS, Brazil\\
   \inst{3}Departamento de Astronomia, Instituto de Física, Universidade Federal do Rio Grande do Sul, CP 15051, 91501-970 Porto Alegre, RS, Brazil\\
             }

   \date{Received 6 June 2024 / Accepted 18 November 2025}

  \abstract{Increasing evidence shows that active galactic nuclei (AGNs) with radio detections have more perturbed ionized gas kinematics and higher outflow detection rates, suggesting a link between radio emission and these processes. In galaxies with weak or ambiguous AGN signatures, some studies attribute the radio emission to star formation, while others propose AGN-driven winds or weak, unresolved jets as the dominant mechanism. To investigate this connection, we take a step back and analyze a sample of star-forming (SF) galaxies with no clear current AGN signatures. Using low-frequency (LOFAR, 144 MHz) and high-frequency (FIRST, 1.4 GHz) radio surveys, combined with spatially resolved spectroscopy from the SDSS-IV MaNGA survey, we compare SF galaxies with 144 MHz detections that either do or do not have gigahertz detections. Despite being matched in stellar mass, redshift, and radio (megahertz) luminosity, gigahertz-detected SF galaxies systematically differ from their non-GHz-detected counterparts. The former display enhanced ionized gas-emission line widths, higher central outflow fractions, redder colors, increased central obscuration, and offset emission-line ratios that shift towards (or closer to) the AGN regime (in the [\ion{N}{ii}]~BPT diagram). Furthermore, the non-gigahertz galaxies are likely undetected due to their extended radio morphologies, while the gigahertz-detected ones are significantly more radio compact. Most of the properties from the GHz-detected (compared to non-detected) remarkably resemble the behavior found in many studies of radio-detected AGNs. This suggests that the underlying physical mechanisms shaping GHz-detected SF galaxies' properties are fundamentally similar. This raises intriguing questions about whether some compact SF galaxies represent a precursor phase of AGN evolution or a form of low-power AGN activity. The radio compact characteristic sizes of GHz-detected SF galaxies also suggest a connection between AGNs and old starburst galaxies.
 }


\maketitle
%

\section{Introduction}

    Several studies have pointed out that the growth of supermassive black holes (SMBHs) can have an impact on their host galaxies as both evolve through cosmic time \citep[e.g.,][]{Fabian_2012,Kormendy_2013,Heckman_2014}. This growth is fueled by gas accretion from secular processes in galaxies \citep{Alexander_2012}, or angular momentum loss via mergers \citep{Hopkins_2006,Hopkins_2009}. Such accretion is today understood to happen either through a singular, highly significant accretion event or via multiple accretion episodes \citep{Harrison_2023} throughout the galaxy's history. These phases are characterized as active galactic nuclei (AGNs), and their presence can manifest throughout multiple wavelengths \citep[e.g.,][]{Padovani_2017_at_all_wave,Alberts_2020,Lyu_2022}. It is known that different techniques to select AGN (in general, multiwavelength-based) can find galaxies with different AGN populations and host galaxy properties \citep[e.g.,][]{Hickox_2009}, potentially indicating not only different powering mechanisms but also AGN in different evolutionary stages. Additionally, obscuration effects influence AGN selection across different wavelengths, as dust and gas can obscure UV, optical, and even X-ray emission \citep{Hickox_Alexander_2018}.

    Radio observations provide an essential window into AGN activity, given that they operate in a low-optical-depth regime. However, the evidence is hampered by the fact that radio emission in galaxies can come from various complex processes \citep{Condon_1992,Panessa_2019}. Star-forming (SF) galaxies (with no current AGN activity) have a radio continuum dominated by a combination of free-free emission and synchrotron radiation \citep{Condon_2016}. Below 30~GHz, most of this emission comes from synchrotron radiation, and above 30~GHz, the contribution of free-free emission from \ion{H}{ii} regions becomes important. Indeed, at 1~GHz, free-free emission is responsible for about 10\% of the radio continuum \citep{Condon_2016}.
    
    In the local Universe, for example, SF galaxies typically dominate at L$_{\text{1.4~GHz}}<10^{23}$~W~Hz$^{-1}$ \citep{Condon_2002}, and their current radio emission is a good tracer of their star-formation rate (SFR). Most of the synchrotron here comes from relativistic electrons, and candidates for accelerating these particles are Type II supernova remnants. These explosions come from short-lived stars more massive than $\sim8$~$M_\odot$, making the radio continuum a good tracer of recent star formation on SF galaxies.

    On the other hand, galaxies hosting AGNs usually show excess radio emission from what we would expect from pure stellar processes \citep{Zakamska_2014}. A relevant component contributing to radio emission in AGN has been attributed to jet (resolved or unresolved) structures \citep{Padovani_2017_at_all_wave}. Controversially, some of the usually so-called radio-quiet AGN display radio emissions that can also be explained by star-formation processes \citep[see][]{Padovani_2016A}. However, alternative processes that can give rise to radio emission in AGN have been proposed in the literature. For example, radio emission can originate due to shocks driven by outflows \citep[e.g.,][]{Zakamska_2016}. \citet{Alban_2024} suggests that the latter could explain why radio-AGN selection techniques can find more perturbed gas compared to a purely optical AGN selection. They show that radio-selected AGNs have larger ionized gas velocity widths when compared to optically selected AGNs.
    
    Indeed, \citet{Torres-Papaqui_2024} found that SDSS optically selected AGN ($z<0.4$) with radio detections display systematically ionized gas emission line widths than non-radio-detected AGN. This has been explored in \citet{Escott_2025}, even finding that the outflow detection rate is increased on radio-detected ($z<0.8$) AGNs \citep[see also][]{Nandi_2025}. Spanning more moderate redshifts ($z<2.5$), studies on red quasars (when compared to blue-selected ones) have shown that they are more likely to be detected through radio observations, having the enhanced kinematics, and more significant obscuration levels \citep[e.g.,][]{Fawcett_2020,Fawcett_2023,Rivera_2023}. Despite including different redshifts or AGN types, these studies show a clear pattern of the role of radio emission in AGNs. While some of these studies have suggested that such differences in the radio properties might be related to outflow-driven shocks, star formation, or low-power jets \citep[e.g.,][]{Torres-Papaqui_2024,Escott_2025,Nandi_2025}, other studies attribute the differences to an AGN evolutionary phase effect, depending on the studied sample \citep[e.g.,][]{Hickox_2009,Fawcett_2023,Alban_2024,Jin_2024}. Interestingly, comparably significant distinctions between radio-detected and non-radio-detected sources are also observed among non-AGN galaxies (see below).

    \citet{Ivezic_2002} discussed that SDSS galaxies with detected radio emission have differing colors, and therefore, also differing morphologies when compared to the overall SDSS galaxies. However, they discuss that such differences significantly reduce when looking at fixed redshift and absolute magnitudes, suggesting that a selection bias drives the differing properties due to the sensitivity of the survey. In contrast, studies focusing specifically on star-forming (SF) galaxies have found evidence for an intrinsic sub-population among radio-detected SF galaxies. For example, \citet{Hopkins_2003} showed that SF galaxies have redder colors and higher obscuration levels (traced by optical colors and Balmer decrements (BD), respectively), which cannot be explained solely by selection effects or sample contamination. The latter results in SF galaxies having differing colors and morphologies. For example, lower BD are observed on bluer galaxies \citep{Stasinska_2001} and analogously, BD have been shown to increase from late to early type galaxies \citep{Stasinska_2004}. \citet{Hopkins_2003} attribute the differences to a combination of optical sample undersampling redder galaxies, or radio detections undersampling the blue ones, both possibilities being attributed to intrinsic physical differences and the differences in the timescales of such processes. Further evidence was presented in \citet{Afonso_2003}, showing that such obscuration in SF radio-detected galaxies happens at a given SFR compared to optically selected SF galaxies. More recently, \citet{Ahmed_2024} confirm that star-forming galaxies with radio detections have systematically higher obscuration levels (traced by BDs), ruling out radio sensitivity limits as the primary cause. These findings suggest that intrinsic physical differences drive the observed trends rather than selection biases alone.

    In summary, it has been shown that both AGN and non-AGN seem to display a sub-population when radio is detected. Motivated by these systematic findings, we attempt to understand the origin of this behavior in a series of papers taking advantage of spatially resolved spectroscopy. In this first paper, we take a step back and analyze this effect on galaxies where no clear AGN is present. If we assume that the radio emission in SF galaxies arises from star-forming processes and not from nuclear activity (AGN), a radio-detected SF galaxy sample (which lies on the SF main sequence) compared to a non-detected one with similar morphologies and SFR properties should not display substantial differences in other properties of their host galaxies. Therefore, we present a comparative analysis in this sample to explore the properties of SF galaxies with and without radio detections at gigahertz frequencies. Our SF galaxy classification is based on central BPT diagnostics following the criteria of \citet{Alban2023}. Our study combines rich multi-wavelength surveys, spatially resolved spectroscopy and stellar population synthesis models. In this way, we can attack the question of whether simple differences related to star formation processes can explain the excess in the kinematics (or other properties) when radio emission is present. The findings can provide hints into the connection between radio emission in active and inactive galaxies and potentially give clues on the episodic nature of AGN.

    The paper is organized as follows. The data and catalogs used for this study are shown in Section \ref{sec:data}. In Section \ref{sec:sample_and_control}, we define samples of radio-detected and undetected star-forming galaxies, carefully removing non-detection biases. And in Section \ref{sec:analysis}, we show our analysis and results. Discussion and conclusions can be found in the last two sections (Section \ref{sec:discussion} and \ref{sec:conclusions}). Throughout this paper, we have assumed a flat $\Lambda CDM$ cosmology with H$_{0}=72$~km~s$^{-1}$~Mpc$^{-1}$, $\Omega_{\rm{M}}=0.3$, and $\Omega_{\Lambda}=0.7$.

\section{Data and catalogs}
\label{sec:data}
    \subsection{MaNGA}

    Combining multi-wavelength data and integral field spectroscopic (IFS) surveys has proven to be insightful. This section details the sources of these datasets used in our study. For IFS, we use the MaNGA SDSS-IV \citep[DR17;][]{Abdurro_2022}. The survey has observed 10010 unique galaxies ($0.01 < z < 0.15$) covering wavelengths between 3622 to 10354 \r{A} with a spectral resolution of R$\sim$2000. The field-of-view depends on the integral field unit array and varies from 12'' to 32'' in diameter, achieving a median physical resolution of around ~1.37~kpc. MaNGA provides already-reduced spectral cubes obtained by the Data Reduction Pipeline \citep[DRP][]{Law2015}, as well as derived spectroscopic quantities obtained from the Data Analysis Pipeline \citep[DAP;][]{Westfall_2019}. We further combine the latter with many MaNGA-focused value-added catalogs from the literature. This not only broadens our parameter space for understanding the properties of our host galaxies, but also shows that the derived properties are consistent between pipelines.

    \subsubsection{Derived values from the MaNGA DRP and DAP}
    \label{sec:manga_DAP}

     We use spectral cubes from the DRP to obtain the [\ion{O}{iii}]~5007 ionized gas kinematics from the per-spaxel spectroscopy. Details of the latter can be found in \citet{Alban_2024}, where we compute the $W_{80}$ maps and effective radius ($R_{eff}$) normalized radial profiles (at annuli steps of 0.25~$R_{eff}$). For emission-line profiles where one or more Gaussians were fitted, the $W_{80}$ measures the width enclosing 80\% of the total flux to account for potential non-gravitational motions. For a single Gaussian profile, $W_{80}\approx2.56\sigma$. We will mainly discuss about the $W_{80}$ from [\ion{O}{iii}]~5007, unless specified otherwise. We use the $W_{80}$ values with a $S/N>3$, and note that our results for $W_{80}$ hold if the $S/N$ cut is set to 7 or 10. We keep this $S/N$ threshold to be consistent with the one applied to the emission lines (see below).
     
     We use the Dn4000\footnote{In this paper, we use Dn4000, which is similar to the standard D4000 but with a narrower measurement band (see \url{https://www.sdss4.org/dr17/manga/manga-analysis-pipeline/}). For simplicity, we will refer to it as D4000.} from the DAP maps and compute their radial profiles the same way as in the $W_{80}$. The D4000 break has been widely understood as a tracer of the mean age of stellar populations. Both D4000 and $W_{80}$ have been corrected for instrumental broadening following MaNGA's documentation guidelines. We get the emission-line ratio DAP-maps ([\ion{O}{iii}]/H$\beta$, [\ion{N}{ii}]/H$\alpha$, [\ion{S}{ii}]/H$\alpha$, [\ion{O}{i}]/H$\alpha$); note that when referring to [\ion{S}{ii}], we are adding both emission lines from [\ion{S}{ii}] (at 6717~\r{A} and 6731~\r{A}). We create maps based on the closest distance to the extreme starburst line \citep{Kewley_2001} from the scatter plot of the [\ion{N}{ii}]~BPT diagram \citep[see][]{kewley_2006}. From all these maps, we also create radial profiles. The properties mentioned above are used (either as resolved maps or integrated annulus) after masking for a signal-to-noise ratio ($S/N$) larger than 3 at each emission line. Constraining this cut to galaxies with a $S/N>10$ does not change our results significantly. In this context, we note that star-forming galaxies are seen to be less affected in comparison to other BPT-classified groups \citep[see][]{Brinchmann2004}.

     \subsubsection{Global, morphological, and environmental quantities}
    
    Global parameters of each galaxy were taken from the value-added catalog from \citet{Sanchez_2022}, which are a product of the PIPE3D software \cite[see][]{Sanchez_2016,Lacerda_2022} applied on the MaNGA survey. The latter includes adopted redshift measurements from the NASA-Sloan Atlas \citep[NSA;][]{Blanton_2011} catalog\footnote{\url{http://www.nsatlas.org}}. Below, we list the parameters using the same key defined in the catalog. The individual definitions following the link in the footnote\footnote{Pipe3D catalog for SDSS-DR17: \url{https://data.sdss.org/datamodel/files/MANGA_PIPE3D/MANGADRP_VER/PIPE3D_VER/SDSS17Pipe3D.html}} and more details can be found in references therein. From this catalog, we use the log\_Mass, log\_SFR\_ssp, T90, D4000\_Re\_fit1, D4000\_alpha\_fit, Re\_kpc, ellip, PA. Namely, stellar mass, star formation rate from H$\alpha$, look-back time at which a galaxy formed 90\% of its current mass, the D4000 stellar index at 1 effective radius and the slope of its gradient (alpha\_fit), and the last parameters are used to normalize the step of the radial profiles and follow the inclination and ellipticity of the galaxy \citep[as described in][]{Alban_2024}.

    Morphological parameters for each galaxy are taken mainly from \citet{Vazquez_2022} and V\'azquez-Mata in prep., an exhaustive visual classification focused on MaNGA galaxies. They perform a visual inspection using the image post-processing and image residuals (r-band) combining the ones from SDSS and the Dark Energy Spectroscopy Instrument \citep[DESI,][]{Dey_2019}. With its deeper imaging compared to SDSS, DESI enables a more refined classification, particularly improving the identification of morphological features in edge-on galaxies. From the catalog, we extract the values of T-Types, which assign a number to the Hubble classification; UNSURE, which flags galaxies with diffuse or compact morphologies where classification was ambiguous; and BARS, which characterizes the identification of a bar. Concerning the Hubble morphological types, T-types~$<0$ are usually associated with elliptical galaxies,  T-types~$\sim0$ correspond to S0s, and  T-types~$>0$ usually refer to spiral Hubble morphologies \citep{Conselice_2006}.

     Additionally, we use the Galaxy Environment for MaNGA Value Added Catalog (GEMA-VAC\footnote{GEMA-VAC: \url{https://data.sdss.org/datamodel/files/MANGA_GEMA/GEMA_VER/GEMA.html}}) which will be described in Argudo-Fern\'andez et al. (in prep). The catalog provides large-scale environment characterizations within 1~Mpc  using a line-of-sight velocity difference of 500 km~s$^{-1}$ \citep[see][]{Argudo_2015}. We use the projected distances (in kpc) to the 1st and 5th nearest neighbors for each MaNGA galaxy.

\subsubsection{Resolved quantities}
    \label{sec:megacubes}

    Stellar population fits for MaNGA have been performed by adapting the STARLIGHT code \citep{Cid_2005} to work with cube data using the {\sc urutau} code \citep{Mallmann_Riffel_2023}. Known as {\large\textsc{megacubes}}, these data products are derived as outlined in Section 3 of \citet{Riffel_2023} and references therein \citep[e.g.,][]{Riffel_2021}. The data products include per-spaxel values of SFRs, stellar population vectors, stellar extinction, and stellar population age weighted by luminosity. Subsequently, {\large\textsc{megacubes}} offers radial profiles from fitted parameters of these maps\footnote{The radial profiles and per-spaxel parameters can be downloaded or interactively seen in the following link: \url{https://manga.linea.org.br/}}. We list here the parameters that are used for analysis in our study:

    \begin{itemize}
        \item Mage/L: mean age weighted by stellar luminosity.
        \item $A_{\rm V}$ Stellar extinction: V-band parameterized extinction obtained from modeling the stellar continuum.
        \item Star formation rate here is defined as the mass processed into stars over an age interval.
        \item Stellar mass: the stellar population mass.
    \end{itemize}

    We note that the radial profiles from the {\large\textsc{megacubes}} are obtained in steps of 0.5~R$_{eff}$, while our radial profiles for the D4000 and $W_{80}$ use steps of 0.25~R$_{eff}$. However, this should not have a significant effect on the comparisons. The radial profiles from {\large\textsc{megacubes}} also accommodate the galaxy's $b/a$ axis-ratio and position angle. The mean or integrated values at each annulus are obtained after masking values where the continuum's $S/N$ is lower than 10. Lastly, in figures, we present the radial profiles of galaxy subsamples by showing the median profile within each annulus as a solid line, with shaded regions indicating the 25th and 75th percentiles.

\begin{figure}
	\includegraphics[width=\columnwidth]{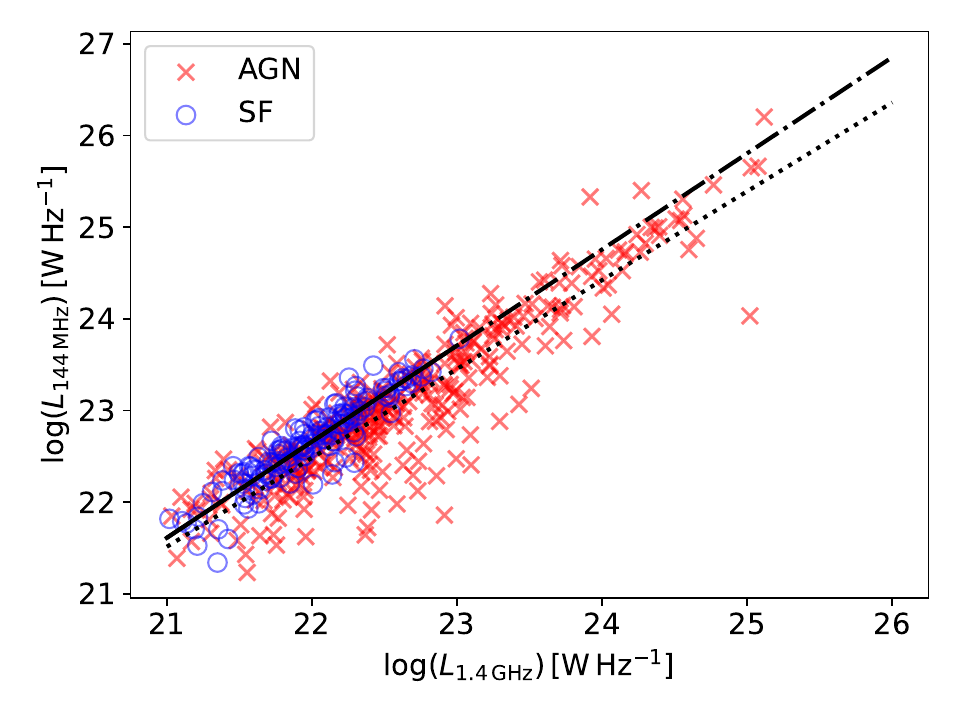}
    \caption{Comparison of the radio continuum luminosities between megahertz and gigahertz frequencies. Blue circles represent SF galaxies and red X-symbols represent AGN candidates chosen from multi-wavelength selection techniques (all the ones mentioned in Section \ref{sec:All_AGN}). The black solid line is a 1D-polynomial fitted to SF galaxies, and it is followed by a dashed-dotted line that extrapolates it to $L_{1.4GHz}>23$~[W~Hz$^{-1}$]. The dotted line is the same fit for AGN candidates.}
        \label{fig:SF_MHz_vs_GHz}
\end{figure}

\subsection{Crossmatching MaNGA with different surveys}

We crossmatch multi-wavelength data from different surveys with MaNGA. All the surveys mentioned here have full coverage of MaNGA galaxies, except for the radio low-frequency data (see below).

\subsubsection{The radio counterparts of MaNGA galaxies}
\label{radio_selection_section}
To gain insights into the origin of the radio continuum of our targets, we combine low and high-frequency radio surveys. We use the Low-Frequency Array \citep[LOFAR,][]{van_Haarlem_2013} two-metre Sky Survey \citep[LoTSS,][]{Shimwell_2017, Shimwell_2022} to obtain the continuum counterparts at 144~MHz and the Faint Images of the Radio Sky at Twenty centimeters \citep[FIRST,][]{Becker_1995} for 1.4~GHz observations. These radio surveys have a comparable resolution, with 6.0'' for LoTSS and 5.4'' for FIRST. While the sky coverage overlap between FIRST and MaNGA is almost a $100$\%, for LOFAR (DR2), the current data release covers around $\sim$60\% of MaNGA.

Several studies have found that radio-detected galaxies form a subpopulation relative to non-detected ones. This has been shown for AGNs \citep[e.g.,][]{Fawcett_2020,Torres-Papaqui_2024,Escott_2025,Nandi_2025}, and for star-forming galaxies \citep[e.g.,][]{Hopkins_2003_b,Afonso_2003,Ahmed_2024}. Given that this behavior has been primarily tested with single-fiber spectroscopic data, we are interested in exploring it by exploiting MaNGA's spatially resolved spectroscopy. In this paper, we start with a sample of SF galaxies to compare sources with and without radio detections. We define non-detections based on the absence of gigahertz continuum emission in the FIRST survey. While we define detections by the presence of GHz, all our galaxies (detected in gigahertz or not) have MHz detections. This allows us to reliably control radio emission distribution when comparing samples with each other. The crossmatch method in \citet{Kukreti_2025} follows standard radio crossmatching techniques commonly used in the literature \citep[e.g.,][]{Best_2012}. The current selection technique ensures that crossmatched galaxies have robust detections ($S/N > 3$) well above the survey’s sensitivity limit. The crossmatch is performed within a 6.0'' radius, and for FIRST, a flux density threshold of 0.5~mJy is applied. Therefore, we emphasize that a gigahertz non-detection does not mean that the source is radio silent at this frequency.

We crossmatch these surveys to MaNGA from \citep{Kukreti_2025} to obtain the luminosities at both frequencies (gigahertz and megahertz) and deconvolved size estimates for the MHz continuum based on the semi-major fitted 2D Gaussian for the cross-matched counterpart, normalized by the galaxy's effective radius. A detailed discussion on the difficulties and artifacts about the size measurements of radio sources can be found in \citet{Shimwell_2022}, where they suggest a criterion to distinguish resolved and unresolved sources. We use the latter to characterize the fraction of unresolved galaxies in our samples.

\subsubsection{The 1.4~GHz vs. 144~MHz relation in SF galaxies}
\label{sec:mhz_vs_ghz}

In the context of SF or non-active galaxies, radio emission is a reliable estimator of recent star formation as it is, in principle, extinction-free \citep{Condon_1992}. Below $\sim$30~GHz, most of the radio continuum is dominated by synchrotron emission. Consequently, their radio continuum in different windows is ubiquitously shown to be correlated with their star formation rate  \citep[SFR; see][for a review]{Kennicutt_2012}. For example, high frequency surveys such as FIRST (at 1.4~GHz) have been extensively used and shown to correlate with the SFR \citep[measured by many indicators; e.g.,][]{Davies_2017,Kennicutt_2009}. Similarly, \citet{Gurkan_2018} have shown the same for lower-frequency surveys such as LoTSS (able to probe at $\sim$144~MHz). Therefore, it is well expected that both bands offer an extinction-free SFR indicator for SF galaxies. Naturally, the emission at low- and high-frequency is expected to be correlated in SF galaxies. Indeed, we show this in Figure \ref{fig:SF_MHz_vs_GHz} (see below).

To select SF galaxies, we use the catalog from \cite{Alban2023}, which, in summary, are BPT-selected (at an aperture of 2~kpc; see the details in the reference). In Figure \ref{fig:SF_MHz_vs_GHz}, we show that such correlation holds with a small scatter for these SF galaxies (highlighted in blue circles). However, the red crosses show a significant scatter for AGN candidates (selected by different techniques; see Section \ref{sec:All_AGN}). We fit 1-D polynomials for SF galaxies and AGN candidates separately (we extrapolate the one from SF galaxies in the plot). This shows that AGN candidates are generally offset from this correlation, suggesting that their radio emission is not solely attributable to SF processes, and likely includes contributions from AGN activity \citep[see][]{Padovani_2016A}. By leveraging this tight relationship for SF galaxies, we can effectively mitigate possible biases between gigahertz detections and non-detections by comparing samples within matched MHz luminosity bins. This is incorporated when constructing our control sample (see the details in \ref{sec:sample_and_control}).

\begin{figure*}

	\includegraphics[width=\textwidth]{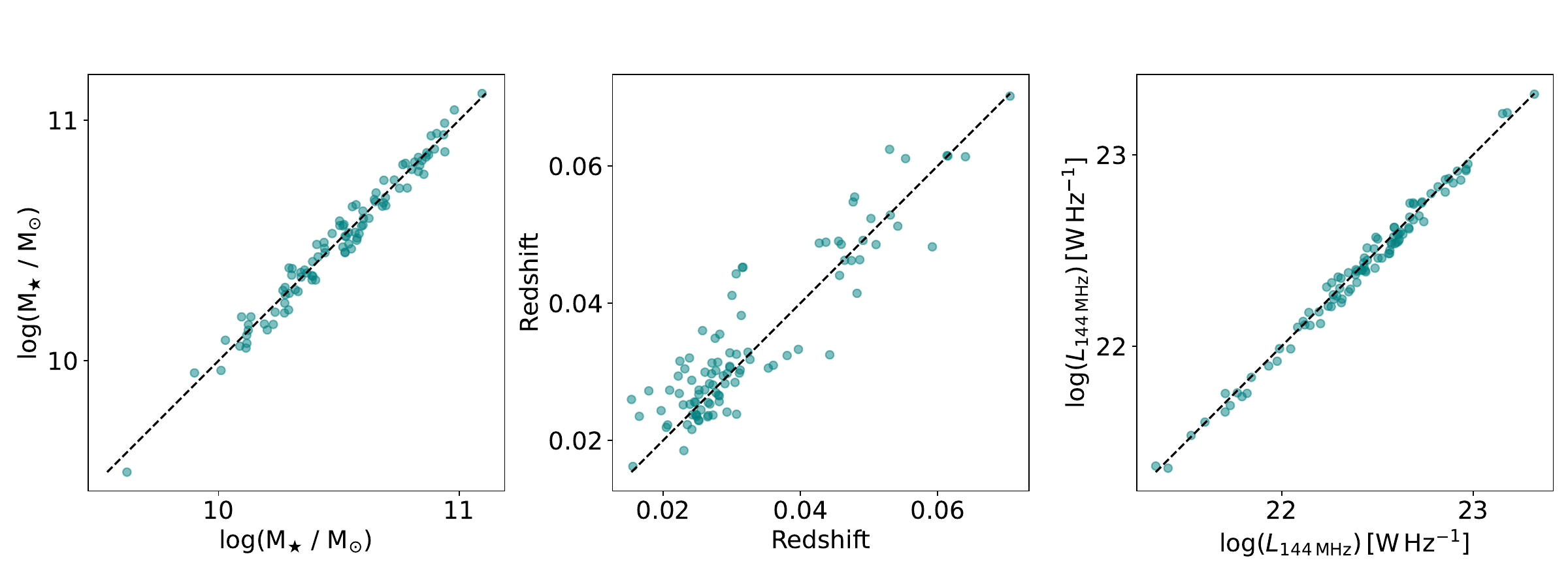}
    \caption{Parameter space used to match pairs of SF galaxies with and without gigahertz detections. We compared the properties of each pair, with GHz-SF on the x-axis and nGHz-SF on the y-axis, where both axes represent the same parameter. The dashed line in each plot represents the 1-to-1 correspondence where the parameter pairs are equal.}
        \label{fig:GHz-vs-nonGHz_prop}
\end{figure*}

\subsubsection{Mid-infrared photometry catalogs}

We cross-match MaNGA galaxies with the Wide-field Infrared Survey Explorer \citep[WISE,][]{Wright_2010}. WISE is a mid-infrared all-sky survey that observes through filters in four different bands: W1 ($3.4\mu m$), W2 ($4.6\mu m$), W3 ($12\mu m$), W4 ($22\mu m$). We use the coordinates from MaNGA sources to find crossmatching targets from the AllWise data release using a 6'' aperture. Mid-infrared luminosities, as obtained by the WISE bands, have been shown to correlate with SFRs for SF galaxies \citep[e.g.,][]{Wen_2013}. In our analysis, we use W1, W2, and W3 with $S/N > 3$. Nevertheless, all the galaxies that we use (see Section \ref{sec:sample_and_control}) have a $S/N$ above this threshold, noting that their spatial resolution is comparable to that of LOFAR and FIRST.

\subsection{MaNGA AGN catalogs used in this work}
\label{sec:All_AGN}

In this section, we list some of the existing MaNGA AGN catalogs we know about up to date. We list each selection technique and direct the reader to the respective references for further details. Catalogs not mentioned may be redundant to the scope of this paper, given the catalogs already listed below. We use the following catalogs:

\begin{itemize}
    \item Mid-infrared colors: \citet{Comerford_2024}.
    \item Broad Balmer lines: \citet{Fu_2023} and \citet{Negus_2024}.
    \item Mid-infrared variability: \citet{Pai_2024}.
    \item X-rays: \citet{Molina_2023} and \citet{Comerford_2024}.
    \item Radio (MHz):  \citet{Kukreti_2025}.
    \item Optically (BPT): \citet{Alban2023}.
    \item Radio (GHz): \citet{Alban_2024}.

\end{itemize}

Approximately 18\% of MaNGA galaxies are classified as AGN in one way or another. While a detailed exploration of their combined relevance might deserve a separate study, it lies outside the scope of the current work. Here, we simply note that only about 30 of our SF galaxies are present in any of the above-mentioned AGN catalogs, with roughly half showing radio-gigahertz detections. Including or excluding these targets does not significantly affect any results discussed in this paper. We emphasize this to clarify that our conclusions are not subject to the presence of current AGN activity unless a population of heavily obscured AGN (see discussion) at low redshifts has managed to evade several AGN multi-wavelength selection techniques.

\begin{figure*}
	\includegraphics[width=\textwidth]{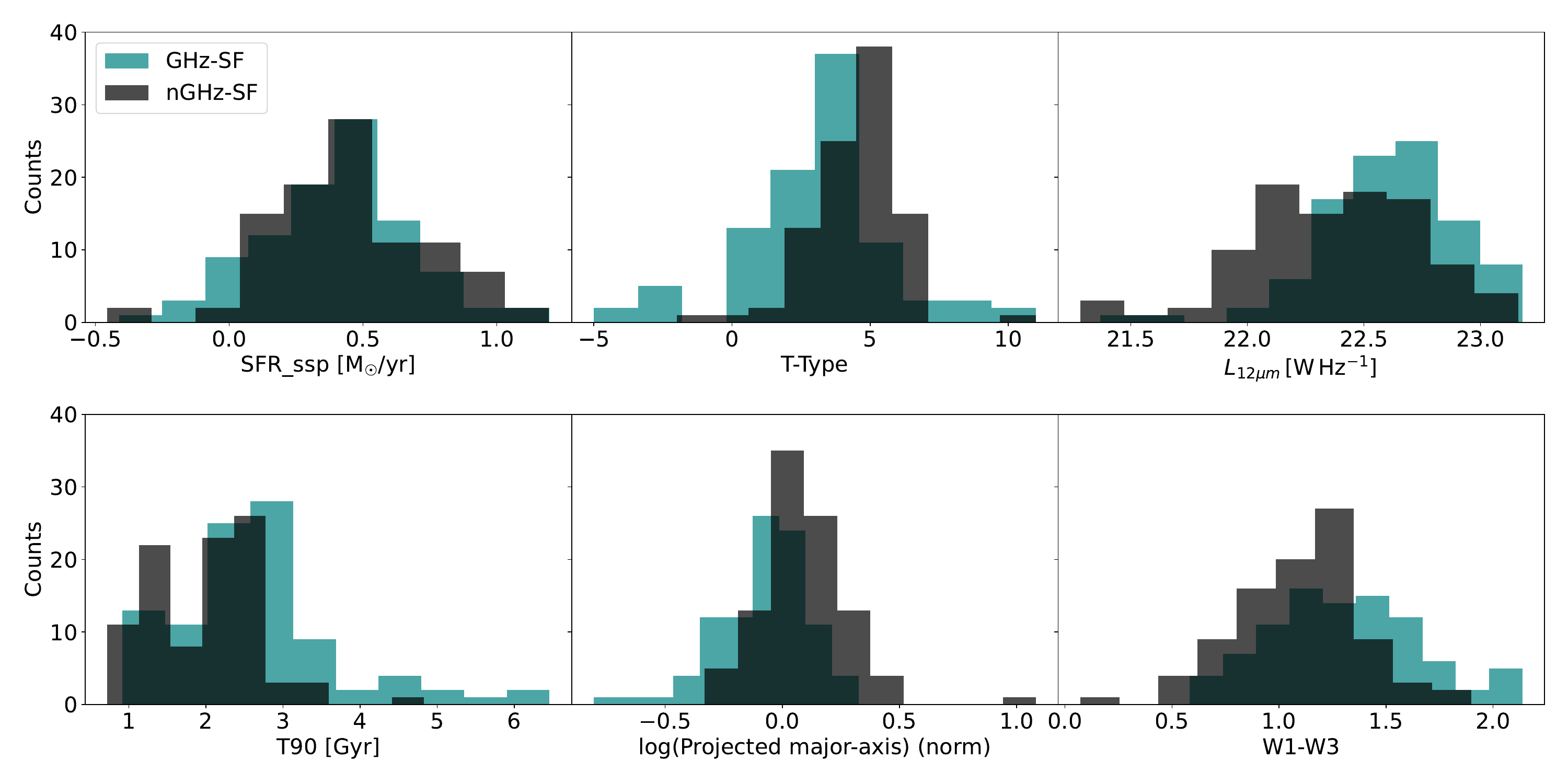}
    \caption{Uncontrolled global or integrated properties of SF galaxies
with and without gigahertz detections.}
        \label{fig:GHz-vs-nonGHz_props_uncontrolled}
\end{figure*}

\section{Characteristics of our sample and their controls}
\label{sec:sample_and_control}

    We aim to explore the origin of radio emission in galaxies by selecting a sample of main-sequence SF galaxies where no clear current AGN is present. Therefore, when referring to SF galaxies in this paper, we have removed from them the AGN mentioned in Section \ref{sec:All_AGN}. We start with a sample of purely SF galaxies (see Section \ref{sec:mhz_vs_ghz}) with 144~MHz detections in LOFAR. This sample is divided into two groups, based on whether a FIRST (1.4~GHz) counterpart was detected or not. While the distinctions reported in Section \ref{sec:mhz_vs_ghz} were initially established using gigahertz detections alone, we include LOFAR due to its greater sensitivity relative to FIRST \citep{Best_2023}. This addition accounts for the fact that star-forming galaxies are expected to produce radio emissions, ensuring that non-detections at gigahertz frequencies are not solely due to sensitivity limitations.
    
    Radio detections, or, in other words, the crossmatch between radio and MaNGA \citep[see Section \ref{radio_selection_section}, and also][]{Kukreti_2025} is defined following general selection techniques used in the literature \citep[see also,][]{Best_2005}. We define gigahertz detected SF galaxies as GHz-SFs, with 139 sources, and non-gigahertz detected SF galaxies as nGHz-SFs, with 813 sources (note that these numbers are after selecting sources with a LoTSS detection). To understand whether the detection of radio emission (specifically gigahertz) impacts other galaxy properties, we create a control sample over a fixed parameter space. For each GHz-SF, we select one nGHz-SF if possible, controlling simultaneously for three parameters: L$_{\text{144~MHz}}$, stellar mass (from the VACs), and redshift.

    Matching in stellar mass and redshift ensures that more massive and nearby galaxies, which are intrinsically easier to detect, are appropriately paired. More importantly, L$_{\text{144~MHz}}$ does not only attempt to pair galaxies by similar extinction-free SFRs but also intends to control for the gigahertz global luminosity that we cannot detect. The latter is based on the tight relation seen between radio emission and the SFR in galaxies (see Section \ref{sec:mhz_vs_ghz}). Figure \ref{fig:GHz-vs-nonGHz_prop} shows the control parameter space comparison between GHz-SFs and nGHz-SFs. We could similarly do this using SFR estimated from H$\alpha$, given the correlation between L$_{\text{1.4~GHz}}$ and SFR(H$\alpha$) in star-forming galaxies. However, as we will show later, the SFR (from H$\alpha$ or the stellar populations) gets self-controlled with L$_{\text{144~MHz}}$ (see Figure \ref{fig:GHz-vs-nonGHz_props_uncontrolled}). This leaves us with a sample of 97 galaxy pairs, removing some GHz-SFs due to a lack of suitable controls in nGHz-SFs. These removed GHz-SF galaxies do not affect the overall results of our study. Including them within a more flexible control sample selection would still yield comparable distributions of control properties between nGHz-SFs and GHz-SFs, albeit at the cost of compromising the pairing of one-to-one galaxy comparisons. Therefore, we adopt a stricter control sample criteria to ensure that variations in behavior are not driven by significant disparities in control parameters (even when the overall distributions would appear similar). During the discussions, we will mainly refer to these (GHz-SFs and nGHz-SFs) galaxies.

\section{Analysis}
\label{sec:analysis}

The strategy of our control sample ensures that GHz-SFs and nGHz-SFs have the same SFR and stellar masses. This section shows how uncontrolled parameters behave in each group when the only difference is whether a galaxy has a gigahertz detection. Measurements of the median values of the global and resolved properties, as well as the p-values from a two-sample Kolmogorov-Smirnov (KS) test, have been compiled in the Appendix (see Table \ref{tab:hists}, Table \ref{tab:radial_4steps} and Table \ref{tab:radial_8steps}).

\subsection{Comparison of integrated or global properties}
\label{sec:int_global_prop}

\begin{figure*}
    \includegraphics[width=\textwidth]{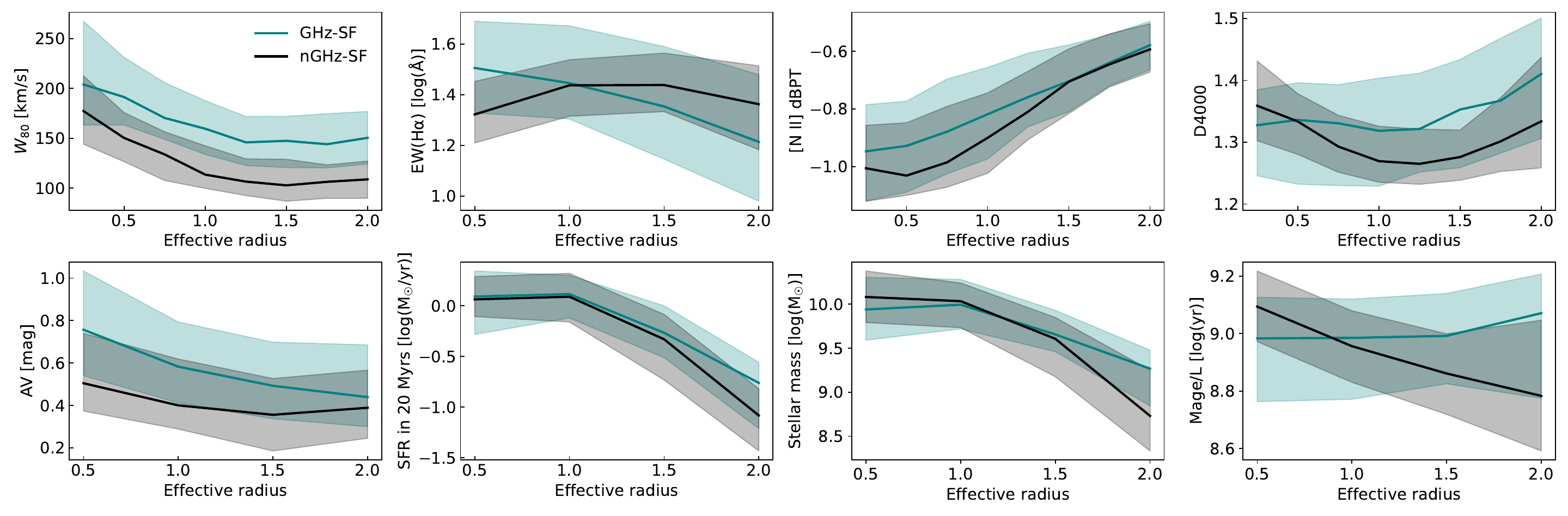}
    \caption{Comparison of modeled and empirically resolved properties between GHz-SFs and nGHz-SFs. The solid lines represent the median value at each annulus and the shaded region represents the 25th and 75th percentiles. The empirical properties are located in the top row, and the modeled, in the bottom row.}
        \label{fig:GHz-vs-nonGHz_prop_more_empirical}
\end{figure*}

In Figure \ref{fig:GHz-vs-nonGHz_props_uncontrolled}, we display a set of parameters not used during the control algorithm. Our sample of GHz-SF and nGHz-SFs galaxies has not been matched directly to have the same SFR; however, as we see in this Figure, our samples have very similar values (we show here the SFRs derived from stellar population modeling of the VACs). Similar results are found when looking at the SFR derived from H$\alpha$ from the VACs \citep[as well as the SFRs derived from][]{Riffel_2021}. The latter can be naturally explained noting that, for star-forming galaxies, there is a correlation between SFR and L$_{\text{144~MHz}}$, and SFR and L$_{\text{1.4~GHz}}$ (see discussion in Section \ref{sec:mhz_vs_ghz}). Furthermore, our galaxies lie on the star-forming main sequence (SFMS), where stellar mass and SFR are tightly correlated. As a result, they occupy a similar locus in the SFMS. We show, therefore, that the global SFR is a parameter that is self-controlled by our control sample and suggests that differences in properties in these galaxies should be independent of SFR.

We use the T-types to infer a morphological classification of these galaxies. Slightly lower T-type values are found on GHz-SFs despite being in similar positions on the main sequence compared to nGHz-SFs (with a mean value of 4.5 in nGHz-SFs and 3.0 in GHz-SFs). This is confirmed by the T-Types and the \textit{best\_n\_type} from the VACs, which directly classifies the galaxy according to its Hubble type. Lower T-types are usually associated with redder colors \citep{Conselice_2006}, which can explain the larger luminosity in WISE filters and the redder WISE colors when compared to nGHz-SFs (however, we still observe this at fixed morphologies).

Earlier spiral types have been associated with the so-called red spirals. Indeed, using $(g-r)>0.63-0.02(Mr+20)$ as a threshold for the latter \citep[see][]{Masters_2010}, shows that $\sim$31\% of our GHz-SFs can be classified as red spirals, while only $\sim$9\% of nGHz-SFs satisfy the condition (if we remove edge-on galaxies, with $log(a/b)<0.2$, to minimize the effect of dust reddening, this would change those numbers to 12\% and 4\% respectively; see Appendix \ref{sec:restricted_sample}). One of the suggestions is that these galaxies are a more evolved version of the usual blue spirals. However, removing the galaxies that satisfy the red spiral threshold does not modify the general differences between GHz-SFs and nGHz-SFs. If red spirals represent an older version of blue spirals, GHz-SFs (even when removing the red spirals) may represent an intermediate population bridging the two. To get insights into the latter in more detail, we show the T$90$ measurements from the VACs, which measure the look-back time at which a galaxy assembled 90\% of its mass. Figure \ref{fig:GHz-vs-nonGHz_props_uncontrolled} shows that, indeed, GHz-SFs are more likely to have assembled their mass earlier than nGHz-SFs.

    Redder colors in GHz-SFs can also indicate higher levels of obscuration and the presence of dust. In SF galaxies, mid-infrared luminosities have been seen to correlate with SFR. Given that our galaxies have the same SFR, this immediately means that a SFR excess is measured from SFR(L$_{\text{W4}}$) or SFR(L$_{\text{W3}}$) for GHz-SFs. Offsets from the latter correlation have been attributed to the presence of an AGN. However, the overwhelming AGN selection techniques we have gathered (see Section \ref{sec:All_AGN}) do not show evidence of any current activity in these galaxies. The excess in infrared colors can also be explained in terms of excess in the kinematics. \citet{Baron_2019} has shown that optically-selected AGN with an outflow detection (which, in our context, translates into finding higher $W_{80}$) have redder colors than those without outflows. The authors interpret the latter as follows, in the sense that outﬂows, common in AGNs, can carry dust which is heated by the AGN emission, reemitting in the infrared. We argue that such an effect could still be present when the optical signature from the AGN has turned off, given that our GHz-SFs resemble this kinematic excess and the corresponding redder colors. Therefore, redder colors in GHz-SFs can be related to pre-heated dust carried by an outflow ejected from a previous AGN event.

    We have analyzed further morphological and environmental characteristics. For example, differences in the observed properties between GHz-SFs and nGHz-SFs do not depend on the environment or whether there's a bar. Using the distance to the 5th nearest neighbor, less than 2\% of galaxies (in GHz-SFs and nGHz-SFs) have more than 5 companions in the surroundings of 200~kpc. If we observe broader scales, no substantial differences arise. Similarly, no differences are seen if looking at the distance to their 1st nearest neighbor. We, therefore, do not expect to have environmental effects impacting our results. As for the radio morphologies (central-bottom panel in Figure \ref{fig:GHz-vs-nonGHz_props_uncontrolled}), the semi-major axis, normalized by the galaxy's effective radius, shows that GHz-SFs are more compact. We note that the unnormalized semi-major axis sizes (in arcseconds) from the GHz-SFs are also smaller than the ones from nGHz-SFs $\sim96$\% of the cases. In this context, using a criterion for distinguishing between resolved and unresolved sources \citep[following the methods from][; see Section 3]{Shimwell_2022}, we find that $\sim$75\% of GHz-SFs are unresolved. In contrast, only $\sim$19\% nGHz-SFs are unresolved. This is true at fixed optically derived properties (see Section \ref{sec:restricted_sample}).

    It can be argued that supernova remnants can also carry outflows and perhaps mimic the observed contrasting properties. However, we find no evidence of an excess of SNRs when using SNR diagnostics in our samples. This is somewhat expected given that the SFR of our targets is similar. We emphasize that in the sample of our current study, we have no AGN active in our targets. An alternative explanation is that given that redder colors can, in turn, translate into stronger star-formation events. However, the latter needs to be true at fixed extinction-free SFR(L$_{\text{144~MHz}}$). We emphasize that the parameters discussed here are model-independent and use a purely empirical comparison between the samples.

Using a two-sample KS test, in Appendix \ref{appendix_statistical_significance} we report the statistical p-values of each parameter from Figure \ref{fig:GHz-vs-nonGHz_props_uncontrolled}. This is shown on a logarithmic scale in Figure \ref{fig:pvalues_hists}. For example, the p-value for SFR yields a value above the typical significance threshold, indicating that we do not statistically find evidence for a difference in the SFR distributions (GHz-SFs versus nGHz-SFs). We also report the p-values for the distance to the close neighbors and the b/a axis ratio, where the same conclusion can be drawn. However, the remaining parameters present low p-values, rejecting the null hypothesis (see Appendix \ref{appendix_statistical_significance}), consistent with the differences reported in this section. Below, we discuss resolved properties to get insights into this observed behavior.

\subsection{Resolved ionized gas kinematics}

The excess in the emission line width is discussed first, as it was the initial parameter we found to differ in the presence of detected radio emission, and it is particularly explored in recent literature. Across all radial bins, GHz-SFs display systematically higher $W_{80}$ values than nGHz-SFs. Despite having similar global and resolved SFRs, the broader emission-line profiles suggest that the process responsible for the broadening may be connected to the compact radio emission morphology. Interestingly, although we are working with SF galaxies, studies on AGN samples have suggested that shocks generated along the outflows can accelerate particles, giving origin to radio emission \citep[e.g.,][]{Zakamska_2016}.

In a sample of radio-detected-AGNs and non-radio-detected-AGNs, \citet{Escott_2025} find that the former are more likely to present outflows and larger $W_{80}$ values. Our targets show no signs of current nuclear activity yet display kinematic excesses resembling the behavior of the AGNs from \citet{Escott_2025}. We further confirm that the kinematic differences persist in spatially resolved spectra. Escott et al. discuss that radio emission is unlikely to originate from high-powered radio jets in their sample. They suggest that other likely origins can be related to low-powered radio jets, AGN outflow-driven shocks, or star formation. Given that our sample has virtually the same SFRs, as shown by many tracers (even resolved), we suggest that star formation processes are unlikely to explain the kinematic differences (although contrasting SF histories might be an alternative explanation; see the discussion in Section \ref{sec:discussion}). Similar results have been found in \citet{Nandi_2025} for radio-detected AGNs. The interpretation of our results becomes challenging given that there is also no clear evidence for a current AGN present today in any of our galaxies. Below, we expand our study to other resolved quantities to gain insights into the possible processes causing this behavior.

\subsection{Modeled and empirical resolved properties}

We split the resolved properties into empirical in the top row (e.g., the D4000 stellar index) and modeled in the bottom row (e.g., stellar extinction) of Figure \ref{fig:GHz-vs-nonGHz_prop_more_empirical}. Empirically, we can observe that when compared to nGHz-SFs, the sample of GHz-SFs (aside from differences in the velocity widths) has a larger EW(H$\alpha$) in the central regions but lower in the outskirts, off-nuclear emission-line ratios closer to the extreme starburst line of the [\ion{N}{ii}] BPT-diagram, and lower D4000 in the inner regions but larger at the outskirts. We emphasize that most radial distribution changes are likely lost or diluted in single-fiber spectroscopic studies, highlighting the importance of spatially resolved samples.

Continuing the comparison between the global stellar mass and star formation rates, we show that the same is true for their resolved behavior (see the bottom row of Figure \ref{fig:GHz-vs-nonGHz_prop_more_empirical}). \citet{Riffel_2021} have compared SFR estimates from the H$\alpha$ emission with SFR estimates obtained from stellar population synthesis models for MaNGA galaxies \citep[see also][]{Mellos_2024}. They show that the SFR over the last 20~Myrs shows the best correlation when compared to other age bins (e.g., SFR over the last 100~Myrs). Therefore, in Figure \ref{fig:GHz-vs-nonGHz_prop_more_empirical}, we have used the SFR averaged over the last 20~Myrs, and similar results are found when we look at the SFR over the last 100~Myrs. We note that some discrepancies can appear at R$_{eff}>1.5$, however, MaNGA covers in general up to 1.5~R$_{eff}$ (in Section \ref{sec:restricted_sample}, we discuss that our results are not dramatically changed by controlling for optical sizes). The relation between SFR and stellar mass for SFMS galaxies has also been shown to hold in resolved data \citep{Sanchez_2020}. Indeed, our GHz-SFs and nGHz-SFs display the same resolved specific SFR (sSFR).

The $A_{\rm V}$ stellar extinction is more elevated in the central regions. The same behavior is observed when looking at the Balmer decrement using H$\alpha$/H$\beta$ (e.g., from the VACs). The luminosity-weighted age follows a similar pattern to that described by D4000. Indeed, the D4000\_Re\_fit (from the VACs), which shows the slope from a 1D-polynomial from the D4000 radial profile, shows that the D4000 has positive slopes for GHz-SFs and more negative for nGHz-SFs. Both D4000 (tracing stellar population ages) and M$_{\text{age}}$/L show that GHz-SFs are slightly younger in the central regions and older in the outskirts when compared to nGHz-SFs. If we explain this behavior only in terms of star-formation processes, this shows that GHz-SFs have slightly younger central stellar populations at fixed SFR and stellar mass compared to nGHz-SFs. Judging by the increased stellar extinction and Balmer decrements in the central regions, an obscured SF event can be in place in a group of GHz-SFs that was supposed to have older stellar populations, inverting their D4000 gradients. However, the latter is unlikely, as we have similar SFRs in both samples.

\begin{figure*}
	\includegraphics[width=522pt]{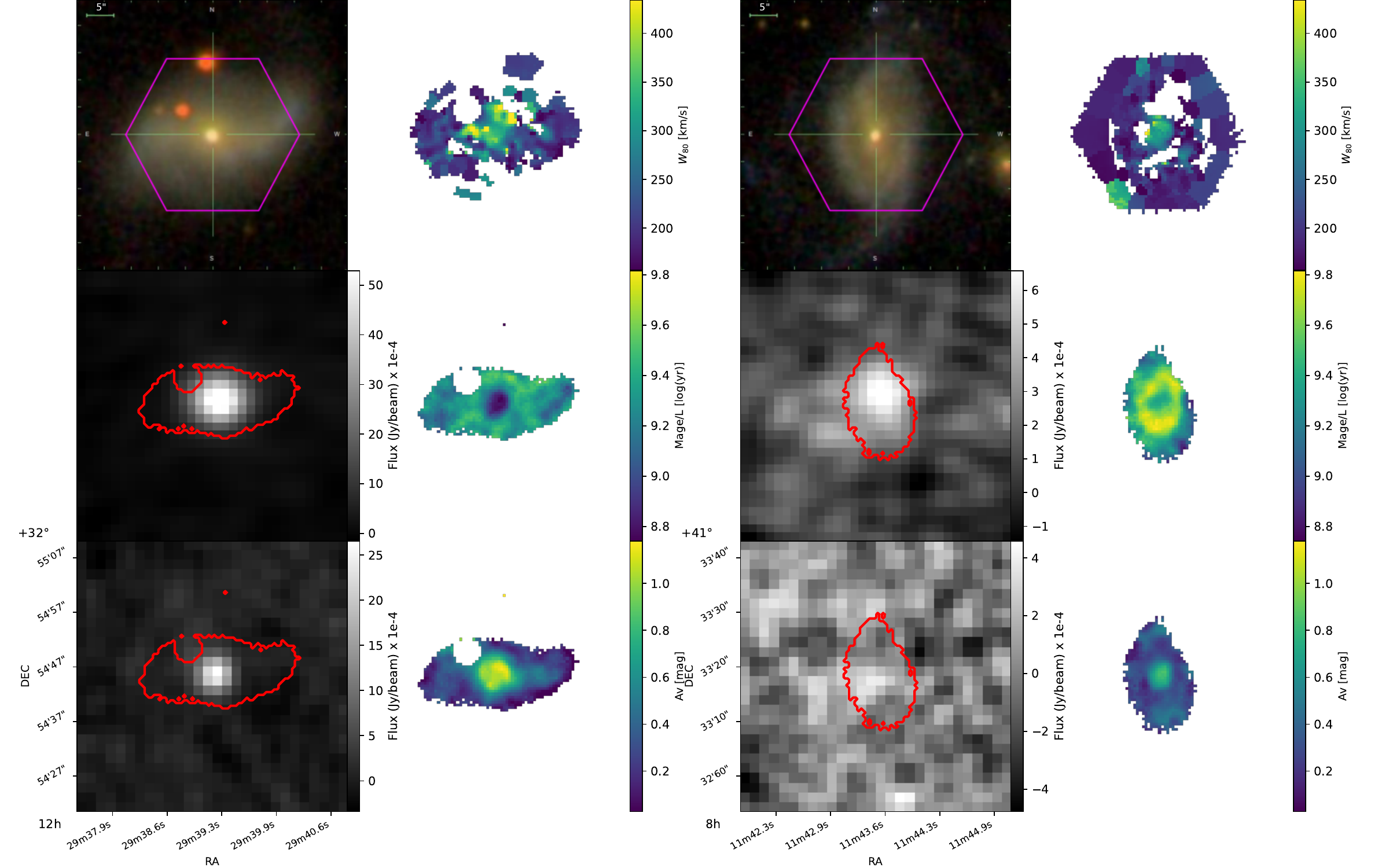}
    \caption{Continuum and spatially resolved properties from a pair of galaxies of the main sample. We show four columns of plots, where the two left columns correspond to GHz-SF and the right ones to nGHz-SF. The GHz-SF and nGHz-SF shown here are twins (control sample pairs) in the parameter space defined in Section \ref{sec:sample_and_control}. The left column in each shows the SDSS optical image at the top, the LOFAR sky cut in the same region as shown in SDSS is seen in the middle, and the FIRST counterpart at the bottom. The right columns show MaNGA's spatially resolved properties. At the top, the W$80$, at the middle, the age-weighted luminosity, and at the bottom, the Visual extinction. The red regions in the radio continuum images display the region of MaNGA's coverage.}
        \label{fig:some_maps}
\end{figure*}

In Figure \ref{fig:some_maps}, we illustrate key properties of a GHz-SF in the two rightmost columns and its nGHz-SFs twin in the two leftmost columns. From top to bottom, the left column in each pair shows the SDSS optical image, LOFAR 144~MHz continuum image, and FIRST 1.4~GHz continuum image. Analogously, the right column presents the $W_{80}$, stellar population age, and stellar extinction (from our measurements and values from the {\large\textsc{megacubes}}; see Section \ref{sec:megacubes}). A quick comparison of the right columns reveals that the resolved trends observed as radial profiles in Figures \ref{fig:GHz-vs-nonGHz_prop_more_empirical} are also evident at the spaxel level. Namely, GHz-SFs are more centrally obscured and seem to have slightly younger stellar populations. Despite having similar MHz radio luminosities, nGHz-SFs likely remain undetected at gigahertz frequencies due to their more extended radio morphologies, also seen statistically in the bottom-left panel of Figure \ref{fig:GHz-vs-nonGHz_props_uncontrolled}. In fact, nGHz-SFs will insist on being more extended than GHz-SFs even in a restricted control sample (see Figure \ref{fig:some_maps2} and Section \ref{sec:restricted_sample}). Notably, FIRST’s detection limit (1~mJy) falls below the contrast scale used for GHz-SFs in both figures. This is likely related to the effect described in \citet{Becker_1995}, where sources at fixed integrated flux densities have a critical size limit above which are likely undetected.

As shown in Section \ref{sec:int_global_prop}, in Appendix \ref{appendix_statistical_significance} we report the p-values from the two-sample Kolmogorov-Smirnov test for the properties shown in Figure \ref{fig:GHz-vs-nonGHz_prop_more_empirical}. Given that these properties are displayed radially, in Figure \ref{fig:pvalues_radial} we report their p-values in the same way, at each corresponding annulus. Notably, as an example, the p-values do not indicate differences in the SFR at any annulus, contrary to the $W_{80}$.

\subsection{Emission-line diagnostics}
\label{sec:em_line_diagnostics}

\begin{figure}

	\includegraphics[width=\columnwidth]{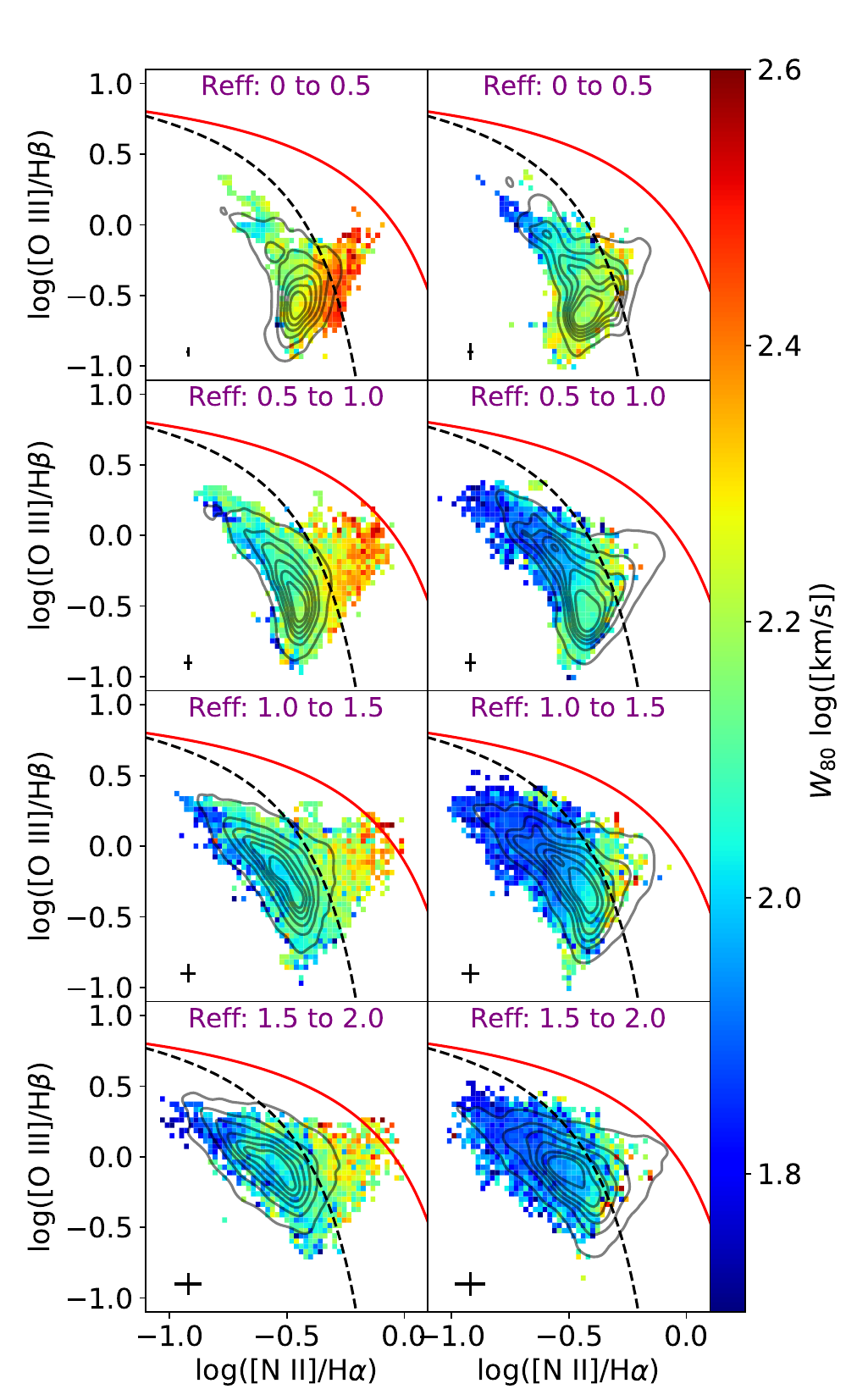}
    \caption{Resolved emission line ratios binned by their $W_{80}$. The first column of plots shows the positions of the spaxels (from the DAP, see Section \ref{sec:manga_DAP}) binned by $W_{80}$ for GHz-SFs. The second column shows the same for nGHz-SFs. The distribution of the emission-line ratio spaxels of either GHz-SF or nGHz-SFs is shown in the opposite columns using black contours (see \ref{sec:em_line_diagnostics}). For example, in the left column, $W_{80}$ is binned for GHz-SF, but the contours show the spaxel distribution of nGHz-SF. In all plots, the \citet{Kewley_2001} and \citet{Kauffmann2003a} lines for transition objects and extreme starbursts are shown in the red solid and black dashed lines, respectively. Bins where fewer than three scatter points were present have been excluded. On the bottom left of each plot, we display representative error bars (in black) for the individual scatter points that compose the shown binning.}
        \label{fig:bpt_contour}
\end{figure}

In Figure \ref{fig:bpt_contour}, we show the resolved distribution of the emission-line diagnostics of the [\ion{N}{ii}] BPT diagram. We use the emission line ratios from the DAP (see Section \ref{sec:manga_DAP}). Each panel in the figure shows spaxels from either the GHz-SF sample (left column) or the nGHz-SF sample (right column), separated by radial annuli in steps of 0.5~R$_\mathrm{eff}$ (per row). The spaxels are plotted according to their emission-line ratios and are color-coded by the average $W_{80}$ in that bin. To facilitate comparison, the distribution of spaxel densities for one sample is overlaid as black contours in the opposite column. For example, in the left column (GHz-SF), spaxels are shown in color according to their $W_{80}$ values. In the corresponding right column (nGHz-SF), black contours indicate the density distribution of those same GHz-SF spaxels. In Appendix \ref{sec:representative_spectra}, we show representative spectra from the emission-lines in play.

The GHz-SFs are more likely to present values in the transition zone and closer to the extreme starburst line. The latter is more evident when looking at regions offset from the center. This suggests that GHz-SF may have had a previous active nucleus, which left AGN-like emission. Variations in the [\ion{N}{ii}]/H$\alpha$ have been routinely related to changes in the metallicity (estimated from the stellar populations, weighted by luminosity). We argue that at fixed spatially resolved metallicity, [\ion{N}{ii}]/H$\alpha$ is still in excess. We note that the kinematic differences and emission-line ratio excesses peak in off-nuclear regions. We argue that this can be related to light echoes of previous AGNs \citep[e.g.,][]{Wylezalek2020}.

We emphasize that we mainly show the behavior of the optical diagnostics from [\ion{N}{ii}]/H$\alpha$, because the comparison of [\ion{S}{ii}]/H$\alpha$ and [\ion{O}{i}]/H$\alpha$ in both samples is broadly similar (see also the discussion in Section \ref{sec_discussion_sforming}). Therefore, unlike [\ion{N}{ii}]/H$\alpha$, we don’t expect clear discrepancies in [\ion{S}{ii}]/H$\alpha$ nor in [\ion{O}{i}]/H$\alpha$ (moreover, [\ion{N}{ii}] has very low S/N in some targets, due to its weaker nature.

Overall, when no clear AGN is present, the presence of radio emission (in the gigahertz continuum) or its compact radio morphology seems to favor slightly earlier T-Type morphologies, which in turn have redder infrared colors. Lastly, many systematics have been addressed in the literature regarding the effects of the $b/a$ axis ratio when studying galaxy properties. In Appendix \ref{sec:restricted_sample}, we show a restrictive sample where this is accounted for. We show that most of the trends discussed persist.

\section{Discussion}
\label{sec:discussion}

Our analysis of star-forming galaxies distinguishes two groups: those with gigahertz detections (GHz-SFs) and those without (nGHz-SFs). Both have been matched in stellar mass, redshift, and $L_{\rm 144MHz}$ (also as a proxy for extinction‐free SFR). At the same time, both groups display similar signatures associated with star formation (global and resolved) and occupy similar space in the main sequence. However, GHz-SFs show distinct properties, including enhanced ionized gas velocity dispersion and off-nuclear emission line ratios, suggesting an additional process at play. Moreover, GHz-SFs are more radio compact (from MHz) than nGHz-SFs. The latter can be true even if matching galaxies for their visual extension or concentration (see Appendix \ref{sec:restricted_sample}).

Our results indicate a strong connection between the presence and compactness of radio emission and the host galaxy properties. In this section, we also show that the fraction of outflows is enhanced in the central regions from GHz-SFs. Given that no differences in SFR are evident, we suggest that GHz-SFs are consistent with radio emission originating in weak shocks (and we cannot discard unresolved weak jets), which persist more effectively in gas that has experienced more significant kinematic perturbation, as seen in our evidence. Nevertheless, if shocks were to be the only source responsible for this difference, our results suggest that star formation (or at least, the recent one) is not the primary factor producing them. Here, we explore and discuss the possible mechanisms behind this behavior.

\subsection{Connection to AGN}

We have shown that GHz-SFs have emission-line ratios closer to the extreme starburst line from \citet{Kewley_2001} in the [\ion{N}{ii}] BPT diagram. Figure \ref{fig:bpt_contour} shows that this is particularly evident when looking at off-nuclear regions with values exceeding the \citet{Kauffmann2003a} line. The emission line ratios from GHz-SFs inside this transition zone also have larger $W_{80}$ values. Nevertheless, enhanced $W_{80}$ values also occur below the \citet{Kauffmann2003a} demarcation line compared to nGHz-SF; a behavior seen at all annuli. \citet{Kewley_2013} have shown that galaxies, even with slow shocks ($\sim100$~km~s$^{-1}$ to 200~km~s$^{-1}$), can mimic the emission line ratios seen in the transition zone. Therefore, the observed off-nuclear AGN-like emission in GHz-SFs may not uniquely point to AGN photoionization. Supernova-driven outflows and minor galaxy interaction shocks could contribute to the enhancement of the observed line ratios. However, given the comparable SFRs and environmental properties between GHz-SFs and nGHz-SFs, such a scenario is unlikely to serve as a sole explanation. A detailed disentanglement of shock contributions (AGN-related or not) is beyond the scope of our analysis.

Interestingly, literature has suggested that light echoes of previous AGNs can explain AGN-like emissions from off-nuclear regions \citep[e.g.,][]{Lintott_2009, Wylezalek2020, Xu_2023}. After an activity period has ended, the galaxy's center now lacks its AGN powering mechanism, and off-nuclear regions can still be ionized by the previous activity, slowly diluting in time \citep{Zubovas_2023}. In this context, a past AGN's outflow (or its cumulative effect) can also have perturbed the gas, enhancing the galaxy's emission-line widths. Over the past decade, several studies have suggested that AGN activity in galaxies can be recurrent, with phases of activity turning on and off over time \citep[e.g.,][]{Sebastian_2019,Vaishnav_2023,Wolnik_2024}. Therefore, not only a single epoch but a cumulative AGN event effect can play a significant role in the evolution of their host galaxies \citep{Harrison_2024}. Based on the off-nuclear AGN-like emission in GHz-SFs and the above-mentioned evidence, we suggest that an explanation for these observations is that galaxies in the GHz-SFs sample are more likely to have had one AGN event (or more) in the recent past than nGHz-SFs. If so, we do not expect that the galaxy has had a jetted AGN episode.

Furthermore, GHz-SFs have enhanced $L_{\rm 12\mu m}$ when compared to nGHz-SFs (we show this in Figure \ref{fig:GHz-vs-nonGHz_props_uncontrolled}). Given that our targets have the same radio luminosity (at 144 MHz), the ratio between the radio luminosity and mid-infrared luminosities is lower for GHz-SFs. The latter resembles the behavior of the red quasars from \citet{Klindt_2019}, which also have a larger radio detection fraction. Mid-infrared luminosities can also be used to estimate global SFRs in galaxies. For example, in a sample of star-forming galaxies in \citet{Lee_2013}, the luminosity at 22~$\mu\text{m}$ and 12~$\mu\text{m}$ is correlated with SFR from  H$\alpha$. We have compared the SFR from H$\alpha$ and from $L_{\rm 144MHz}$ with $L_{\rm 12\mu m}$. We observe that GHz-SFs offset from such expected relation, having larger SFRs estimated from infrared filters. Offsets from expected SFR estimators have been commonly associated with starburst galaxies or AGNs. Similarly, redder WISE colors are found in GHz-SFs. The redder WISE colors may explain why GHz-SFs have slightly earlier T-types. Note that, as discussed in Section \ref{sec:analysis}, outflows from AGN in different studies have been seen to carry dust.

 Redder infrared colors in GHz-SFs resemble the behavior seen in AGNs (also $z<0.15$) from \citet{Baron_2019}, where they show that larger velocity widths are observed when AGNs have redder colors, and suggest that outflows from AGNs are carrying dust. This has also been seen by studies at larger redshifts \citep[e.g.,][]{Rivera_2021,Stacey_2022}, and supported by numerical models \citep[e.g.,][]{Faucher_2012}, suggesting a connection between the presence of dust and winds. Curiously, the properties of our GHz-SFs resemble some of the characteristics of the so-called red quasars (although we do not directly claim a direct correspondence in their physical origin, but perhaps similar). \citet{Fawcett_2023} has shown that quasars selected by their red colors have higher chances of being radio-detected when compared to their blue counterparts. More strikingly, their results reveal that the radio detection fraction increases in sources with high extinction. The third-left plot of Figure \ref{fig:GHz-vs-nonGHz_prop_more_empirical} reveals that GHz-SFs resemble the same behavior. \citet{Fawcett_2023} have interpreted their findings as red quasars being a transitional phase that will eventually, through outflows or winds, evacuate the dust and gas to reveal a blue quasar. Although their study focuses on higher-redshift sources than those analyzed here, the similarity to GHz-SF objects suggests a similar origin, potentially as less powerful, low-redshift analogues.

\begin{figure}

	\includegraphics[width=\columnwidth]{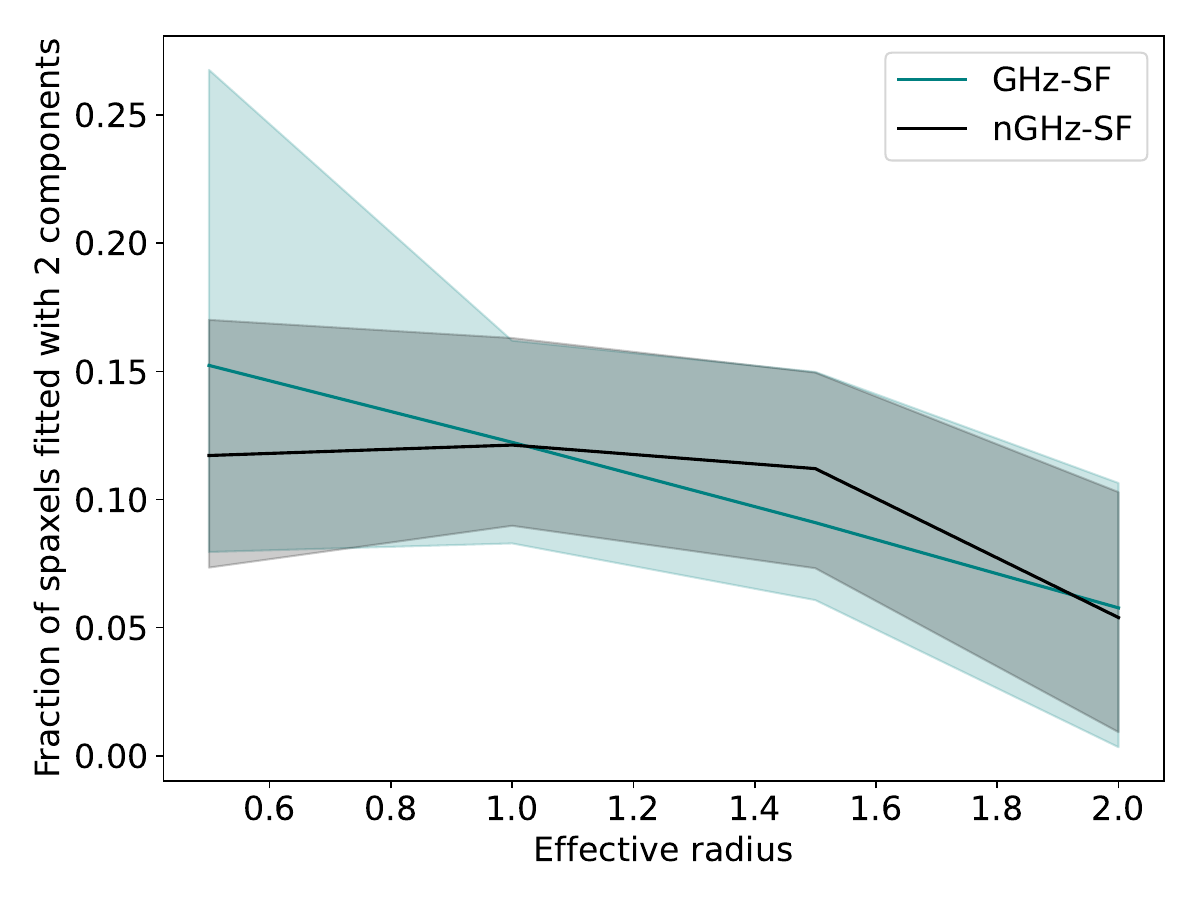}
    \caption{Radial outflow fraction. The plot shows the median fraction of pixels that, at each annulus, were fitted with more than 2 Gaussian components, suggesting an outflow.}
        \label{fig:out_frac}
\end{figure}

We note that the kinematic enhancements in GHz-SFs strikingly mirror the behavior of AGN when they are radio-detected \citep[e.g.,][]{Torres-Papaqui_2024, Escott_2025}, also independent of their star formation. \citet{Escott_2025} find that AGN with radio detections have higher chances of presenting outflow signatures. We perform a similar exercise for our star-forming galaxies. In Figure \ref{fig:out_frac}, we report the median radial profiles for GHz-SFs and nGHz-SFs based on the fraction of spaxels (in annuli steps of 0.5~R$_{eff}$) where a second component is needed. The decision of whether a second component is needed when fitting the emission line profile is described in \citet{Alban_2024} and is more detailed in their Appendix. In summary, we run two individual fits to the [\ion{O}{iii}]: one with one Gaussian fitting, and the other with two Gaussians. A further step decides whether to keep the second component based on a comparison of their reduced chi-squares. We show an example of our fitting procedure applied to a GHz-SF in Appendix \ref{sec:representative_spectra}. We show that our population of GHz-SFs has a slightly greater chance of fitting an additional Gaussian component than nGHz-SFs when looking at the central regions, suggestive of an increased outflow rate. This excess in non-active galaxies independent of the SFR is consistent with the idea that past AGN episodes (or cumulative weak AGN events) may leave behind “fossil” outflows that persist and drive turbulence long after the central engine has turned off \citep[from 2 to 3 times longer than the AGN event, or even 10 times in extreme cases;][]{Zubovas_2023}.  In this context, galaxies experiencing recent AGN episodes would be more likely to have experienced shock events and, therefore, nGHz-SFs either have yet to undergo an AGN event or have suffered a weaker event, leaving only extended MHz, given that emission at higher frequencies cools earlier than lower frequencies \citep[e.g.,][]{Jiang_2010}. In simulations, it has been found that the cooling timescale of a shocked wind can last several Myrs \citep{Zubovas_2023}, although for a single shock event, the gigahertz emission may fade during the AGN phase, before the shock fully dissipates \citep{Jiang_2010}. 

The similarity in the environments of GHz-SF and nGHz-SF galaxies, characterized by the proximity of the closest neighbors, suggests that environmental factors are unlikely to explain the observed differences. Even if the restricted sample (Section \ref{sec:restricted_sample}) shows a slight close neighbor excess in GHz-SFs, it still demonstrates that it cannot be the dominant factor. Therefore, we suggest that this can be explained by GHz-SFs potentially being composed of galaxies that have experienced more (or more recent) AGN events in the past. If the latter is accurate, we do not claim that a previous AGN was present in all GHz-SFs, but GHz-SFs are more likely to have experienced an AGN (or more than one) than nGHz-SFs. Ultimately, our GHz-SFs are systematically more compact in their 144~MHz radio morphology. In a sample of optically selected AGN with gigahertz emission, \citet{Mullaney_2013} found that [O~III]$\lambda$5007 profiles appear more disturbed when 1.4 GHz radio emission originates from a compact core \citep[see also][]{Molyneux_2019,Miranda-Marques_2025}, supporting a link between radio compactness and kinematic disturbances and they suggest this connection to be a result of young or weak radio jets. We cannot rule out that a very young AGN is present in GHz-SFs. Studies with increased resolution have shown that a significant population of AGN might be hidden in such compact cores \citep[e.g.][]{Morabito_2025}. Part of our GHz-SFs could not only be previous AGN, but also restarting.

In summary, many observed properties in GHz-SFs resemble behaviors from radio-detected AGN, suggesting that similar physical processes may be involved. Specifically, radio-detected AGN have also been seen to present larger outflow rates, larger $W_{80}$ values, and are more likely to have radio compact morphologies together with redder colors. However, a pure AGN explanation cannot account for all the observed behavior in GHz-SFs. The younger stellar populations in the center, as indicated by the D4000 index or by stellar population models, suggest an additional mechanism contributing to the observed properties.

\begin{table*}[h]
    \centering
    \caption{Summary of the global or resolved properties of GHz-SFs when compared to nGHz-SFs.
    }
\label{table: summary}
    \begin{tabular}{llccc}
        \toprule
        \textbf{Empirical properties} & \textbf{Behavior of GHz-SFs compared to nGHz-SFs} & \textbf{Resolved} & \textbf{Global} \\
        \midrule
        Star formation rates & Identical or very similar & \checkmark & \checkmark \\
        \textnormal{[\ion{O}{iii}]} velocity dispersion & Excess at all annuli& \checkmark & \\
        Outflow detection rate & Central and off-nuclear excess & \checkmark &  \\
        \textnormal{[\ion{N}{ii}] BPT} & Central and off-nuclear excess & \checkmark &  \\
        D4000 stellar index & Inverted, slightly younger in the center & \checkmark &  \\
        Redshift or distance to the target & Similar & & \checkmark  \\
        Galaxy morphology & Earlier T-Types & & \checkmark  \\
        Wise colors (W1-W3, W2-W3) & Redder colors & & \checkmark  \\
        144~MHz morphology  & More compact semi-major axis size (norm) & & \checkmark  \\
        5th nearest neighbor & Similar, but slightly closer neighbors & & \checkmark  \\
        \midrule
        \textbf{Modeled properties} &&&\\
        \midrule
        Stellar masses & Identical or very similar & \checkmark & \checkmark \\
        Star formation rates & Identical or very similar & \checkmark & \checkmark \\
        Stellar population ages & Inverted, slightly younger in the center & \checkmark & \\
        Central metallicity & Higher in the central regions & \checkmark &  \\
        Stellar extinction & Excess at all annuli & \checkmark &  \\
        Look-back time of 90\% mass assembly & Earlier, faster & & \checkmark  \\
        
\bottomrule
    \end{tabular}
    \label{tab:GHz-SF_vs_nrgfs}
\end{table*}

\subsection{Star formation processes}
\label{sec_discussion_sforming}

We have shown that the fundamental differences between GHz-SFs and nGHz-SFs remain at fixed SFRs. Given that SNRs are one of the candidates for accelerating particles into synchrotron emission, one possibility is that a recent star-formation event has systematically left a larger population of SNRs in the radio-detected population, provided that nGHz-SFs appear to have different star formation histories (e.g., see the T90s in Figure \ref{fig:GHz-vs-nonGHz_props_uncontrolled}). A commonly used approach to distinguish between typical HII regions and SNRs often relies on [\ion{S}{ii}]/H$\alpha>0.4$ as a threshold for SNRs \citep[e.g.,][]{Matonick_1997,Dodorico_1980}. The latter is based on pioneering work from \citet{Mathewson_1973}, where it was observed that the strength of [\ion{S}{ii}] relative to H$\alpha$ in SNRs is larger than in \ion{H}{ii} regions. These SNRs are thought to have stronger shock fronts compared to \ion{H}{ii} regions, behind which a sufficiently high-density cool region allows collisionally excited ionization states, such as [\ion{S}{ii}]. These studies often rely on either Galactic sources or very nearby galaxies, usually depending on high-resolution data \citep[e.g.,][]{Cid_Fernandes_2021,Li_2024}, which is not available for our sample. However, given that we have IFU data available, if a higher fraction of SNRs is present in GHz-SF galaxies, some excess in the [\ion{S}{ii}]/H$\alpha$ ratio can be expected. For example, \citet{Li_2024} has shown that SNRs in resolved spectroscopy on nearby galaxies can display excess in the emission line ratios, even if SNRs are blended with HII regions. In our study, the [\ion{S}{ii}]/H$\alpha$ radial profiles of both GHz-SFs and nGHz-SFs look almost identical, with GHz-SFs even having slightly smaller values. Nevertheless, only a few sources have values that cross the threshold of [\ion{S}{ii}]/H$\alpha>0.4$ in any of the samples, with 6 galaxies in nGHZ-SF and 1 in GHz-SF.

Other more elaborated diagnostics for SNRs, based on photoionization and shock excitation models, rely on comparing [\ion{S}{ii}]/H$\alpha$ to [\ion{O}{i}]/H$\alpha$ or employing the [\ion{O}{i}]~BPT diagnostic \citep{Kopsacheili_2020}. However, we have done these comparisons, and none of these diagnostics show an excess in favor of a population of SNRs in GHz-SFs. However, the sizes of SNRs can be several orders of magnitude smaller than MaNGA's resolution. \citet{Li_2024} show that blended SNR can mimic HII-emission-like regions. Therefore, this effect may still be present but not observable in our sample.

More flexible diagnostics for SNRs have been tested after correcting blended SNR emission line fluxes to get their intrinsic emission line ratios \citep{Cid_Fernandes_2021}. It is important to note that, as \citet{Cid_Fernandes_2021} shows, the SNR contribution/impact to the global properties of galaxies is negligible in terms of when measuring SFR (estimated from H$\alpha$) or when testing emission-line ratios. For example, \citet{Cid_Fernandes_2021} finds that only 0.7\% of the H$\alpha$ from their studied galaxy contributes to the overall SFR. Similarly, a detailed study in a larger sample shows that, on average, 5\% of the H$\alpha$ flux can come from SNRs \citep{Vucetic_2015}. Conversely, this is not a concern in our study and only shows that we would overestimate their SFR if SNRs were indeed in excess in GHz-SFs. Such an overestimation would bias GHz-SFs to have slightly larger SFRs while keeping lower line widths. If this overestimation were the case, our differences in the kinematics would be even more significant. It is unlikely that a population of SNRs can leave mid-infrared emission imprints in the global properties of our host galaxies. Studies of SNRs in mid-infrared colors show that, even if detected, they are still too faint. For example, \citet{Lee_2005} and more references therein show that in the mid-infrared, only around 16\% of 100 confirmed galactic SNRs were detected at around $\sim4.5-5.8\mu m$. This is expected as the integrated H$\alpha$ luminosity from SNRs is also far from dominant in the global emission in their hosts.

Alternatively, as indicated in the restricted sample, if star formation (SF) were the dominant factor underlying the differences between GHz-SFs and nGHz-SFs, it would imply distinct ongoing SF histories for the two populations. We can get some insights into the latter by looking at the T90s in Figure \ref{fig:GHz-vs-nonGHz_props_uncontrolled}, which suggest that GHz-SFs have evolved faster. Similarly, the radio compact morphologies of GHz-SFs may lead to a fraction of GHz-SFs being older starbursts, presumably in their initial path to quiescence \citep[this has been seen at $0.5<z<3.0$;][]{Gomez_2019}. This, in turn, can be discussed in terms of the AGN-starburst connection \citep{Alexander_2012,Ishibashi_2016}, and has been explored in limited samples. Indeed, an important component in galaxy evolution, potentially driving star formation to quiescence, has been discussed by the role of Post-starburst galaxies \citep[see the comprehensive review from][]{French_2021}. For example, post-starburst galaxies have been seen to present AGN light echoes \citep[in MaNGA galaxies][]{French_2023}, consistent with a delayed AGN effect on the galaxy.

Despite being out of the scope of our study, we have briefly compared our sample (GHz-SFs and nGHz-SFs) with samples of post-starburst (PSB) galaxies for MaNGA. \citet{Cheng_2024} classifies poststarburst galaxies in different categories: irregular, central, and ring-like poststarbursts. A simple look at the gigahertz or non-gigahertz detected poststarbursts (that also have MHz in the LOFAR DR2), shows that around 20\% are gigahertz detected and 80\% are not. For our specific sample and control, we find that 14\% and 19\% of GHz-SFs and nGHz-SFs have a poststarburst classification from \citet{Cheng_2024}. Irregular post starbursts have most of the counts (95\%; only one target here is ring-like PSB) in nGHz-SFs, while 50\% are irregular and the other 50\% are ring-like, in GHz-SFs. \citet{Cheng_2024} discusses that ring PSBs in their sample show positive D4000s and negative EW($H\alpha$). Curiously, GHz-SFs are likely to have the latter behavior, and also more GHz-SFs have ring-like PSB (compared to nGHz-SFs), although this represents a very small fraction of all GHz-SFs. Our results are not changed if we remove all PSBs from MaNGA and repeat the analysis.

Although we have not found evidence of a potential population of post-starburst galaxies in GHz-SFs, the similarity in the SFRs averaged over 20 or 100 Myrs with contrasting stellar population ages suggests that some GHz-SFs could have experienced an earlier (and perhaps modest) star-formation event, rejuvenating their central regions. If so, in the past, this would place the instantaneous (central) SFR of such galaxies in excess when compared to nGHz-SFs. Interestingly, \citet{Gatto_2025} observed that optically-selected AGN have larger central SFRs compared to control galaxies, suggesting a link between AGN activity and star formation. Modern consensus in AGN studies accepts AGN as transient events and cautions against assuming that the impact of AGN in their host galaxies is instantaneous \citep[see a relevant review in][]{Alexander_2025}. For example, \citet{Hickox_2014} have explored a model where all star-forming galaxies host an AGN in averaged timescales of $\sim$100~Myrs. In this model, the average SMBH accretion rate is tightly correlated with the galaxy's SFR when averaged over 100 Myrs, while the AGN accretion rate fluctuates dramatically during this timescale. The model has demonstrated that it produces the observed weak correlations between AGN activity and SFR in AGN studies and reproduces the observed AGN luminosity functions up to redshift 2 \citep[see][]{Hickox_2014}. Therefore, supports a significant evolutionary connection between AGN and their associated star formation processes.

In summary, although star formation processes may explain aspects of our observations, we favor a more integrative picture in which both AGN and star-formation activity are fundamentally linked.

\section{Conclusions}
\label{sec:conclusions}

Radio-detected galaxies have been found to differ when compared to their non-detected counterparts. These differences span from modest to moderate changes, such as colors, morphological types, and obscuration levels in the case of SF galaxies, to complex differences such as strongly differing obscuration and gas velocity dispersions in AGN. This effect has been observed or replicated for AGN and SF galaxies at multiple radio frequencies \citep[e.g., $\sim144$~MHz to 1.4~GHz;][]{Hopkins_2003_b,Ahmed_2024,Torres-Papaqui_2024,Escott_2025,Nandi_2025}. To gain insights on the origin of these findings, we study non-detections based on gigahertz frequencies and virtually follow them up with available and more sensitive low-frequency data. We primarily use SF galaxies to first understand whether the differences can arise without the presence of a currently active AGN.

We have compared star-forming galaxies with (GHz-SF) and without (nGHz-SF) radio-gigahertz detections at fixed stellar masses, 144~MHz luminosities, and redshifts. From a multi-wavelength perspective, none of these galaxies present evidence of having a currently active AGN. Despite showing the same SFRs (estimated by several tracers), both samples systematically differ in global and resolved properties. In Table \ref{table: summary}, we summarize the comparisons we have discussed. Below, we provide an overview of our main results:

\begin{itemize}
    \item[i)] Radio detection in galaxies, whether they are AGNs or not, impacts their properties. We have tested this at 1.4~GHz detections. Analogous findings are seen in other studies at lower or equal frequencies.
    \item[ii)] As summarized in Table \ref{table: summary}, when compared to nGHz-SFs at fixed stellar masses and SFRs, our GHz-SFs are more centrally obscured; they have larger ionized gas velocity dispersions accompanied by larger outflow detection rates. Redder colors and earlier galaxy morphologies. Inverted stellar age gradients while appearing slightly younger in the center, and they have assembled their current stellar mass faster.
    \item[iii)] Broadly speaking, the behavior in GHz-SFs mentioned resembles the behavior from radio-detected AGN of many recent studies. Specifically, larger velocity widths, more obscuration, redder, higher outflow detection rates, and more compact morphologies are seen in GHz-SFs.
    \item[iv)] GHz-SFs have off-nuclear emission line ratios closer to AGN-like signatures. Together with being more radio compact, it appears that GHz-SFs might have suffered an AGN in the recent past, or have an old starburst, probably also related to the connection between starbursts and AGNs.
    \item[v)] SF processes (or at least recent ones) seem unlikely to be the dominant process in explaining all the contrasting behavior. However, an earlier SF event could be responsible for a younger population of stars in the central regions of GHz-SFs. This would agree with a co-evolutionary picture where the same gas reservoir triggers central star formation and AGN activity.
    \item[vi)] Environmental processes and bars present in the morphologies do not seem to have an impact on these differences.
\end{itemize}

    Concerning what is found in the literature, we confirm that for SF galaxies, radio detections come preferentially from galaxies that are more obscured by visual extinction. Broadly suggesting them to be a sub-population of SF galaxies themselves \citep[as seen in][]{Hopkins_2003}. Similar findings have been found for AGN sources. Interestingly, AGNs also have larger values in their velocity widths and are more likely to show outflow detections when radio emission is detected (and/or has compact morphologies), discarding star-formation processes as the main responsible driver. We speculate that the similarity of the behavior in AGN hosts and SF galaxies when radio is present suggests that historic AGN events could partly explain the different behavior seen in GHz-SFs. If this were not the case, the contrasting properties when radio is detected would need to be typical behavior of galaxies regardless of AGN activity.

    Given that SFR tracers do not seem to be different in any of the probed scales (e.g., 20 to 100 Myrs), combined with the persistent younger stellar populations in GHz-SFs may indicate an earlier star-formation event. This view fits into the modern picture where, while star-formation events slowly vary through time (in Gyrs), AGN events can significantly vary on short time scales ($<100$~Myrs). It serves as a possible channel to trace the effect of AGN during their inactive phases. For example, if the case of more recent AGN events in GHz-SF's hosts was in play, the enhanced kinematics in GHz-SFs contain evidence of the impact of fossil outflows \citep[e.g.,][]{Zubovas_2023}.
    
    We show that for a galaxy sample where each target must have a radio detection, a preferred parameter space in a low-redshift sample is already set. Notably, some of the behavior seen in our results can be washed out using single-fiber spectroscopic data, emphasizing the utility and motivation to exploit IFU data. Both points mentioned above should be held in mind when studying high-redshift targets. In a future study, we will repeat this exercise, BPT-transition objects, quenched galaxies, and AGNs.

\begin{acknowledgements}

    M.A. acknowledges Andrew Hopkins for insightful discussion during the initial stages of this work. M.A. also extends gratitude to the GALENA research group for their invaluable discussions, which have significantly shaped the ideas presented in this paper.\\ 

    D.W. acknowledges support through an Emmy Noether Grant of the German Research Foundation, a stipend by the Daimler and Benz Foundation and a Verbundforschung grant by the German Space Agency.\\

    RAR acknowledges the support from Conselho Nacional de Desenvolvimento Cient\'ifico e Tecnol\'ogico (CNPq; Proj. 303450/2022-3, 403398/2023-1, \& 441722/2023-7), Funda\c c\~ao de Amparo \`a pesquisa do Estado do Rio Grande do Sul (FAPERGS; Proj. 21/2551-0002018-0), and CAPES (Proj. 88887.894973/2023-00).\\

    RR acknowledges support from  Conselho Nacional de Desenvolvimento Cient\'{i}fico e Tecnol\'ogico  ( CNPq, Proj. CNPq-445231/2024-6,311223/2020-6,  304927/2017-1, 400352/2016-8, and  404238/2021-1), Funda\c{c}\~ao de amparo \`{a} pesquisa do Rio Grande do Sul (FAPERGS, Proj. 19/1750-2 and 24/2551-0001282-6) and Coordena\c{c}\~ao de Aperfei\c{c}oamento de Pessoal de N\'{i}vel Superior (CAPES, 88881.109987/2025-01).\\

    This project makes use of the MaNGA-Pipe3D dataproducts. We thank the IA-UNAM MaNGA team for creating this catalogue, and the Conacyt Project CB-285080 for supporting them.\\

    Funding for the Sloan Digital Sky Survey IV has been provided by the Alfred P. Sloan Foundation, the U.S. Department of Energy Office of Science, and the Participating Institutions. SDSS-IV acknowledges support and resources from the Center for HighPerformance Computing at the University of Utah. The SDSS web site is www.sdss.org.\\

    SDSS-IV is managed by the Astrophysical Research Consortium for the Participating Institutions of the SDSS Collaboration including the Brazilian Participation Group, the Carnegie Institution for Science, Carnegie Mellon University, the Chilean Participation Group, the French Participation Group, Harvard-Smithsonian Center for Astrophysics, Instituto de Astrof\'isica de Canarias, The Johns Hopkins University, Kavli Institute for the Physics and Mathematics of the Universe (IPMU) / University of Tokyo, the Korean Participation Group, Lawrence Berkeley National Laboratory, Leibniz Institut f\"ur Astrophysik Potsdam (AIP), Max-Planck-Institut f\"ur Astronomie (MPIA Heidelberg), Max-Planck-Institut f\"ur Astrophysik (MPA Garching), Max-Planck-Institut f\"ur Extraterrestrische Physik (MPE), National Astronomical Observatories of China, New Mexico State University, New York University, University of Notre Dame, Observatario Nacional / MCTI, The Ohio State University, Pennsylvania State University, Shanghai Astronomical Observatory, United Kingdom Participation Group, Universidad Nacional Autonoma de M\'exico, University of Arizona, University of Colorado Boulder, University of Oxford, University of Portsmouth, University of Utah, University of Virginia, University
\end{acknowledgements}

\bibliographystyle{aa}
\bibliography{bibliography}

\begin{appendix}

\section{Statistical significance of parameters}
\label{appendix_statistical_significance}

\begin{figure}[b]
	\includegraphics[width=\columnwidth]{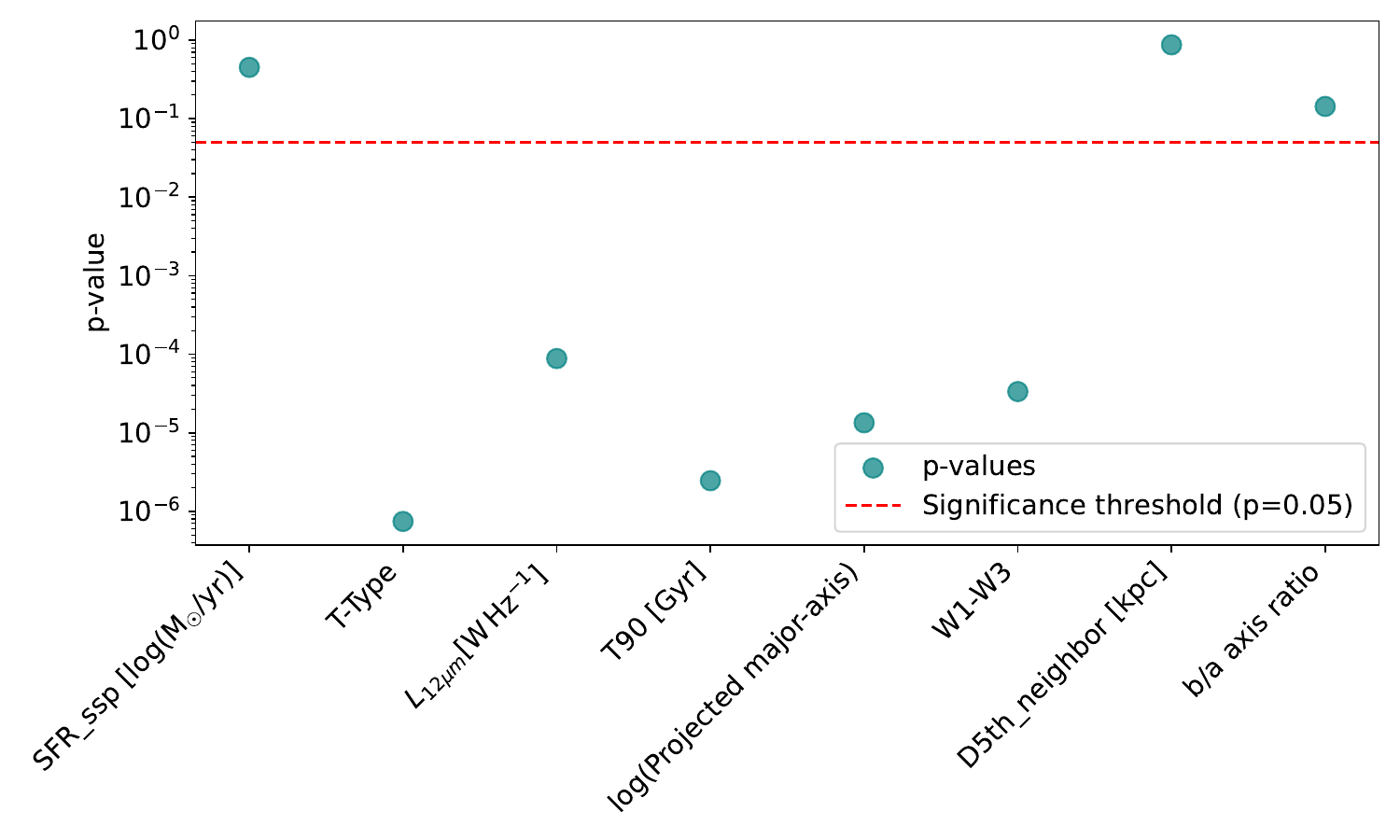}
    \caption{Kolmogorov-Smirnov test for integrated or global properties. The red-dashed line shows $p=0.05$.}
        \label{fig:pvalues_hists}
\end{figure}

\begin{figure*}
\centering
	\includegraphics[width=0.9\textwidth]{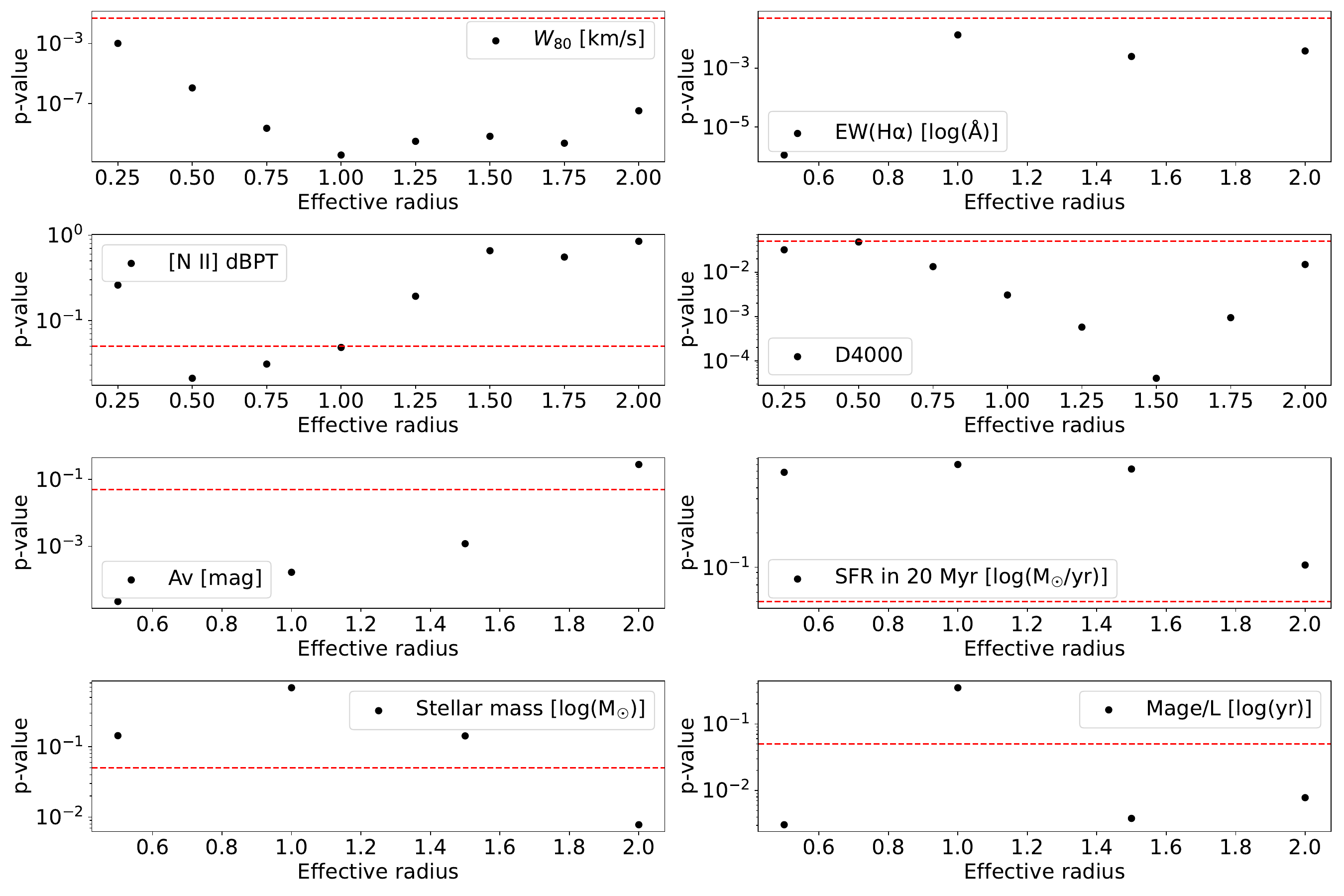}
    \caption{Kolmogorov-Smirnov test for radial profiles. We show the p-values at each annulus for each parameter of Figure \ref{fig:GHz-vs-nonGHz_prop_more_empirical}. The red-dashed lines show $p=0.05$.}
        \label{fig:pvalues_radial}
\end{figure*}

We perform two-sample Kolmogorov-Smirnov tests to obtain the statistical p-values. Low p-values (e.g., $p<0.05$) reject the null hypothesis that the compared samples are related and indicate that the observed differences are unlikely to have occurred by chance. In Figure \ref{fig:pvalues_hists}, we show the p-values for the integrated/global galaxy properties shown in Figure \ref{fig:GHz-vs-nonGHz_props_uncontrolled}, and, additionally, we have added the values for the environmental characteristics and $b/a$ axis ratio. We include the latter two here to show that they also do not exhibit significant differences between GHz-SFs and nGHz-SFs. The p-values at each annulus for the radial profiles are shown in Figure \ref{fig:pvalues_radial}, following the order of the parameters adopted in Figure \ref{fig:GHz-vs-nonGHz_prop_more_empirical}. For the case of the outflow fraction (Figure \ref{fig:out_frac}), the only value with statistical significance (a p-value below 0.05) is in the central annuli. We summarize our results in three tables:

\begin{itemize}
    \item In Table \ref{tab:hists}, the first column lists the parameters shown in Figure \ref{fig:GHz-vs-nonGHz_props_uncontrolled}. The second and third columns give their median values for GHz-SFs and nGHz-SFs, and the last column lists the p-values from the two-sample KS test.

    \item Table \ref{tab:radial_4steps} shows the 25th, 50th (median), and 75th percentiles used for the radial profiles in Figures \ref{fig:GHz-vs-nonGHz_prop_more_empirical} and \ref{fig:out_frac}. This table includes only parameters with radial profiles sampled in 4 annuli.

    \item Table \ref{tab:radial_8steps} follows the format of Table \ref{tab:radial_4steps}, but for parameters with radial profiles sampled in 8 annuli (from Figure \ref{fig:GHz-vs-nonGHz_prop_more_empirical}).
\end{itemize}

\begin{table}[!htbp]
\centering
\caption{Median values from GHz-SFs and nGHz-SFs for parameters reported in Figure \ref{fig:GHz-vs-nonGHz_props_uncontrolled} and their two-sample KS test p-value.}
\label{tab:hists}
\begin{tabular}{cccc}
\toprule
Parameter & GHz-SF & nGHz-SF & log(pval) \\
\midrule
SFR\_ssp & 0.413 & 0.467 & -0.346 \\
T-Type & 3.0 & 5.0 & -6.129 \\
$L_{\rm 12\mu m}$ & 22.633 & 22.406 & -4.054 \\
T90 & 2.569 & 2.094 & -5.610 \\
log(Projected & -0.0450 & 0.0580 & -4.872 \\
 major-axis) (norm) &  &  &  \\
W1-W3 & 1.301 & 1.119 & -4.475 \\
\bottomrule
\end{tabular}
\end{table}

\begin{table*}[b]
\centering
\caption{Summary of the 25th, 50th (the median), and 75th percentiles for each parameter shown in Figure \ref{fig:GHz-vs-nonGHz_prop_more_empirical} that has values averaged in 4 annuli steps and the outflow fraction from Figure \ref{fig:out_frac}, together with their two-sample KS test p-value.}
\label{tab:radial_4steps}
\begin{tabular}{lrrrr}
\toprule
$r/R_{eff} \in$ & $[0.0,0.5]$ & $[0.5,1.0]$ & $[1.0,1.5]$ & $[1.5,2.0]$ \\
\midrule
\textbf{GHz-SFs} & & & & \\
\midrule
EW(H$\alpha$)\_25th & 1.328 & 1.304 & 1.145 & 0.979 \\
EW(H$\alpha$)\_50th & 1.505 & 1.445 & 1.353 & 1.213 \\
EW(H$\alpha$)\_75th & 1.689 & 1.671 & 1.589 & 1.479 \\
AV\_25th & 0.540 & 0.411 & 0.336 & 0.299 \\
AV\_50th & 0.756 & 0.582 & 0.492 & 0.438 \\
AV\_75th & 1.033 & 0.793 & 0.698 & 0.685 \\
SFR(20 Myrs)\_25th & -0.282 & -0.120 & -0.511 & -1.211 \\
SFR(20 Myrs)\_50th & 0.091 & 0.114 & -0.266 & -0.761 \\
SFR(20 Myrs)\_75th & 0.342 & 0.300 & -0.001 & -0.558 \\
Stellar mass\_25th & 9.591 & 9.719 & 9.460 & 8.844 \\
Stellar mass\_50th & 9.940 & 9.994 & 9.656 & 9.268 \\
Stellar mass\_75th & 10.304 & 10.281 & 9.930 & 9.477 \\
Mage/L\_25th & 8.763 & 8.771 & 8.825 & 8.775 \\
Mage/L\_50th & 8.983 & 8.985 & 8.992 & 9.071 \\
Mage/L\_75th & 9.126 & 9.120 & 9.139 & 9.207 \\
Out\_frac\_25th & 0.079 & 0.082 & 0.060 & 0.003 \\
Out\_frac\_50th & 0.152 & 0.122 & 0.091 & 0.057 \\
Out\_frac\_75th & 0.267 & 0.162 & 0.149 & 0.106 \\
\midrule
\textbf{nGHz-SFs} & & & & \\
\midrule
EW(H$\alpha$)\_25th & 1.209 & 1.314 & 1.333 & 1.182 \\
EW(H$\alpha$)\_50th & 1.322 & 1.436 & 1.438 & 1.362 \\
EW(H$\alpha$)\_75th & 1.452 & 1.538 & 1.564 & 1.513 \\
AV\_25th & 0.373 & 0.288 & 0.185 & 0.244 \\
AV\_50th & 0.504 & 0.400 & 0.355 & 0.388 \\
AV\_75th & 0.739 & 0.618 & 0.526 & 0.566 \\
SFR(20 Myrs)\_25th & -0.106 & -0.160 & -0.732 & -1.431 \\
SFR(20 Myrs)\_50th & 0.062 & 0.089 & -0.330 & -1.082 \\
SFR(20 Myrs)\_75th & 0.287 & 0.319 & -0.085 & -0.818 \\
Stellar mass\_25th & 9.790 & 9.733 & 9.175 & 8.332 \\
Stellar mass\_50th & 10.082 & 10.034 & 9.606 & 8.730 \\
Stellar mass\_75th & 10.377 & 10.241 & 9.848 & 9.251 \\
Mage/L\_25th & 8.972 & 8.830 & 8.719 & 8.591 \\
Mage/L\_50th & 9.094 & 8.956 & 8.861 & 8.783 \\
Mage/L\_75th & 9.218 & 9.079 & 8.999 & 9.046 \\
Out\_frac\_25th & 0.073 & 0.089 & 0.073 & 0.009 \\
Out\_frac\_50th & 0.117 & 0.121 & 0.112 & 0.053 \\
Out\_frac\_75th & 0.170 & 0.163 & 0.149 & 0.102 \\
\midrule
\textbf{log(p-values)} & & & & \\
\midrule
EW(H$\alpha$) & -5.952 & -1.873 & -2.667 & -2.246 \\
AV & -4.6468 & -3.773 & -2.740 & -0.426 \\
SFR(20 Myrs)& -0.165 & -0.096 & -0.387 & -2.138 \\
Stellar mass & -0.844 & -0.165 & -0.935 & -3.326 \\
Mage/L & -2.5142 & -0.456 & -2.847 & -1.777 \\
Out\_frac  & -2.318 & -0.353 & -0.585 & -0.003 \\
\bottomrule
\end{tabular}
\end{table*}

\begin{table*}
\centering
\caption{Summary of the 25th, 50th (the median), and 75th percentiles for each parameter shown in Figure \ref{fig:GHz-vs-nonGHz_prop_more_empirical} that has values averaged in 8 annuli steps, together with their two-sample KS test p-value.}
\label{tab:radial_8steps}
    \begin{tabular}{lrrrrrrrr}
\toprule
$r/R_{eff} \in$ & $[0.0,0.25]$ & $[0.25,0.5]$ & $[0.5,0.75]$ & $[0.75,1.0]$ &
$[1.0,1.25]$ & $[1.25,1.5]$ & $[1.5,1.75]$ & $[1.75,2.0]$ \\
\midrule
\textbf{GHz-SFs}&&&&&&&&\\
\midrule
W$_{80}$\_25th & 163.247 & 163.298 & 148.955 & 133.770 & 122.706 & 120.608 & 120.322 & 124.133 \\
W$_{80}$\_50th & 203.975 & 191.372 & 170.374 & 159.565 & 145.893 & 147.432 & 144.028 & 150.433 \\
W$_{80}$\_75th & 267.428 & 231.125 & 205.840 & 187.720 & 171.835 & 171.954 & 174.645 & 176.893 \\
$[$\ion{N}{ii}$]$\_25th & -1.119 & -1.088 & -1.024 & -0.973 & -0.861 & -0.809 & -0.720 & -0.663 \\
$[$\ion{N}{ii}$]$\_50th & -0.947 & -0.928 & -0.878 & -0.818 & -0.758 & -0.704 & -0.640 & -0.579 \\
$[$\ion{N}{ii}$]$\_75th & -0.785 & -0.773 & -0.696 & -0.656 & -0.606 & -0.578 & -0.542 & -0.496 \\
D4000\_25th & 1.245 & 1.232 & 1.229 & 1.228 & 1.251 & 1.258 & 1.282 & 1.306 \\
D4000\_50th & 1.327 & 1.336 & 1.330 & 1.318 & 1.321 & 1.352 & 1.367 & 1.410 \\
D4000\_75th & 1.384 & 1.395 & 1.392 & 1.403 & 1.411 & 1.433 & 1.468 & 1.500 \\
\midrule
\textbf{nGHz-SFs}&&&&&&&&\\
\midrule
W$_{80}$\_25th & 144.064 & 126.519 & 107.612 & 99.705 & 92.532 & 86.875 & 89.789 & 89.855 \\
W$_{80}$\_50th & 177.286 & 150.343 & 133.783 & 113.528 & 106.515 & 102.770 & 106.299 & 108.760 \\
W$_{80}$\_75th & 212.636 & 175.778 & 156.550 & 142.408 & 129.370 & 128.920 & 123.322 & 127.074 \\
$[$\ion{N}{ii}$]$~dBPT \_25th & -1.119 & -1.099 & -1.071 & -1.023 & -0.905 & -0.816 & -0.724 & -0.672 \\
$[$\ion{N}{ii}$]$~dBPT \_50th & -1.005 & -1.031 & -0.985 & -0.900 & -0.809 & -0.705 & -0.645 & -0.593 \\
$[$\ion{N}{ii}$]$~dBPT \_75th & -0.856 & -0.848 & -0.791 & -0.744 & -0.669 & -0.591 & -0.539 & -0.504 \\
D4000\_25th & 1.302 & 1.280 & 1.251 & 1.235 & 1.231 & 1.238 & 1.252 & 1.258 \\
D4000\_50th & 1.358 & 1.333 & 1.292 & 1.269 & 1.264 & 1.275 & 1.301 & 1.333 \\
D4000\_75th & 1.431 & 1.378 & 1.343 & 1.325 & 1.321 & 1.319 & 1.371 & 1.437 \\
\midrule
\textbf{log(p-values)}&&&&&&&&\\
\midrule
W$_{80}$ & -2.9890 & -5.9524 & -8.6419 & -10.4225 & -9.5093 & -9.6440 & -11.5921 & -9.6574 \\
$[$\ion{N}{ii}$]$~dBPT & -0.7238 & -1.6788 & -1.6537 & -1.3173 & -0.6880 & -0.2786 & -0.0998 & -0.0599 \\
D4000 & -1.4934 & -1.3173 & -1.8735 & -2.5142 & -3.2408 & -4.4680 & -2.7811 & -1.7685 \\
\bottomrule
\end{tabular}
\end{table*}

\section{A restricted sample}
\label{sec:restricted_sample}

In Section \ref{sec:sample_and_control}, we establish that GHz-SFs and nGHz-SFs share the same stellar mass, L$_{\text{144~MHz}}$, and redshifts. Here we explore whether the effect of $b/a$ can affect our conclusions. This criterion has been used to minimize the effect of dust reddening \citep[$log(a/b)<0.2$;][]{Masters_2010}. This parameter is not included in the primary analysis for two reasons. First, neither GHz-SFs nor nGHz-SFs show significant differences in $b/a$ (see \ref{fig:pvalues_hists}). This can be expected given their slightly differing morphologies (elliptical galaxies are less common at lower $b/a$, and late-type galaxies tend to exhibit flatter $b/a$ distributions \citep{Tempel_2011, Buitrago_2013, Padilla_2008}). Nevertheless, we account for this parameter when measuring the radial profiles, and its effect should be minimal. Second, we aim to maximize the number of targets without compromising the robustness of the control process. Refining the control sample by including $b/a$ significantly reduces statistical power.

 Lastly, we want to know whether galaxies that look extremely similar (from the point of an analysis of their rest-frame optical characteristics) can still be different because of their radio properties. The motivation to do this is the fact that GHz-SFs would insist on also being optically compact when including control parameters. Therefore, we chose the R$90$ from the VACs as the extent of the optical size (defined as the size at which 90\% of the total V band integrated flux is achieved) and included this parameter in the control sample. We examine whether key trends persist under this stricter selection and if new insights can be learned. The trade-off between sample size and robustness of this galaxy pairing leaves us with 48 galaxy pairs. We anticipate that the results below are very similar if, instead of R$90$, a concentration (C) parameter is chosen or other radius estimations such as the Petrosian radius from the MaNGA-DRPALL. In this comparison, the environment also does not seem to play a significant role in the differences discussed.

Much of what we observe from GHz-SFs and nGHz-SFs persists in the restricted sample. The $W_{80}$ radial profile persists in excess at all annuli for GHz-SFs, and the global SFRs are kept self-controlled. Similarly, the stellar extinction and the Balmer decrement, both tracers of obscuration, remain in excess for the restricted GHz-SFs, although not as drastically as in the main sample. Strikingly, T$90$ values have a more evident excess between both samples. A similar behavior is seen in EW(H$\alpha$) and the T-Type morphological value. Reproducing Figure \ref{fig:bpt_contour} with the restricted sample gives similar behavior to the primary sample. A slight increase in the closest neighbors is observed in GHz-SFs. Indeed, later morphologies are expected to occupy less dense environments \citep{Goto_2003}. An important change to the observed behaviors is the appearance of higher values in the resolved SFRs for GHz-SFs. The latter, even if there is a slight difference, suggests that GHz-SFs have had different star-formation histories than nGHz-SFs. The results here are observed at fixed stellar mass, L$_{\text{144~MHz}}$, redshifts, simultaneously avoiding edge-on galaxies and matching for the visual extension (R$90$). The number of targets in this exercise is reduced to 48 targets, and in turn, we do not conclusively claim that this will always be true at fixed optical extensions. This emphasizes a known issue \citep{Hopkins_2003}, where constraining a sample to only objects with radio detections is subject to the shown behavior. However, this restricted sample comparison broadly recover the property differences from the main sample.

\begin{figure*}[h]
	\includegraphics[width=522pt]{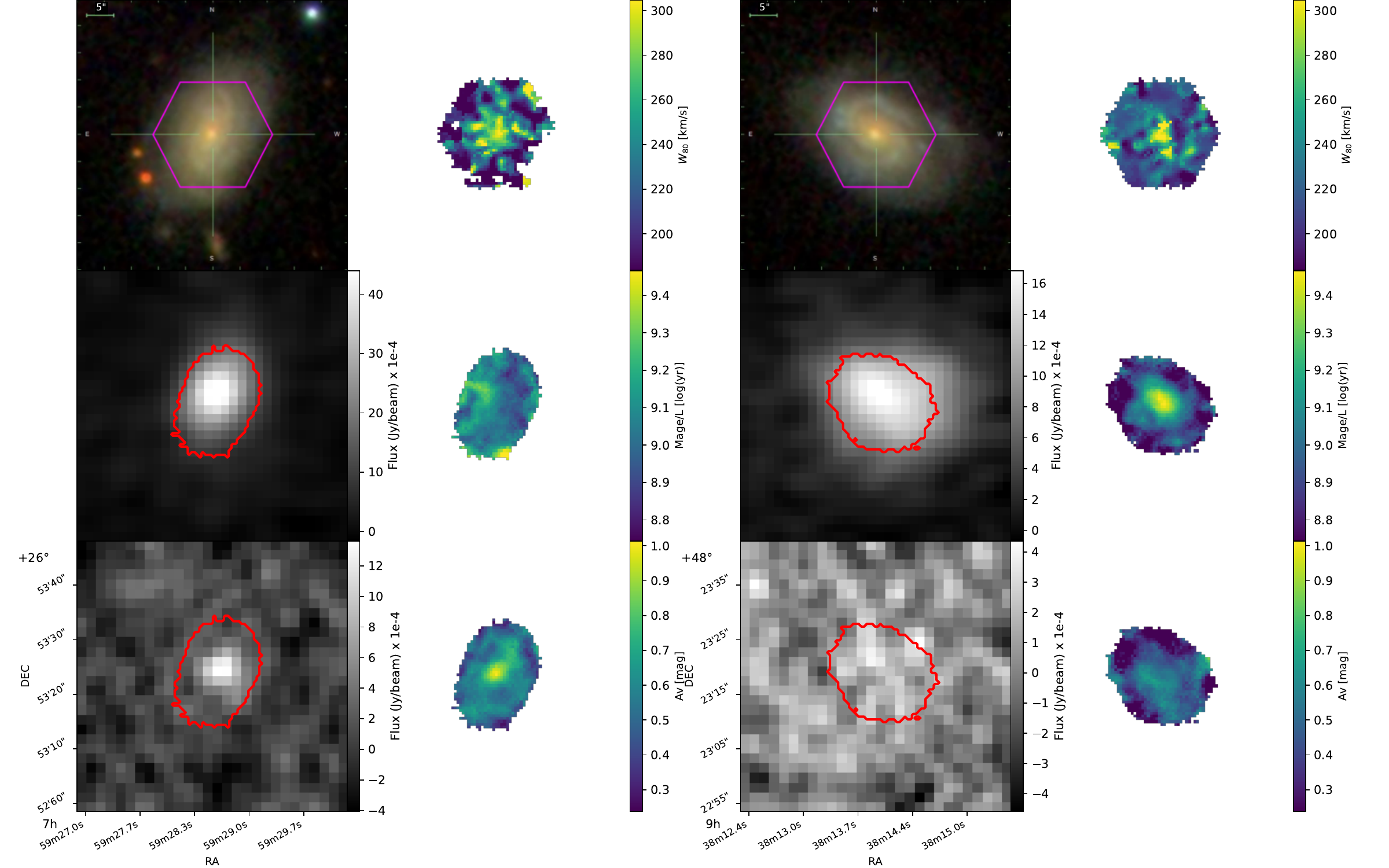}
    \caption{Continuum and spatially resolved properties from a pair of galaxies of the restricted sample. The caption description follows Figure \ref{fig:some_maps}. }
        \label{fig:some_maps2}
\end{figure*}

\section{Representative spectra.}
\label{sec:representative_spectra}

\begin{figure*}[h]
	\includegraphics[width=522pt]{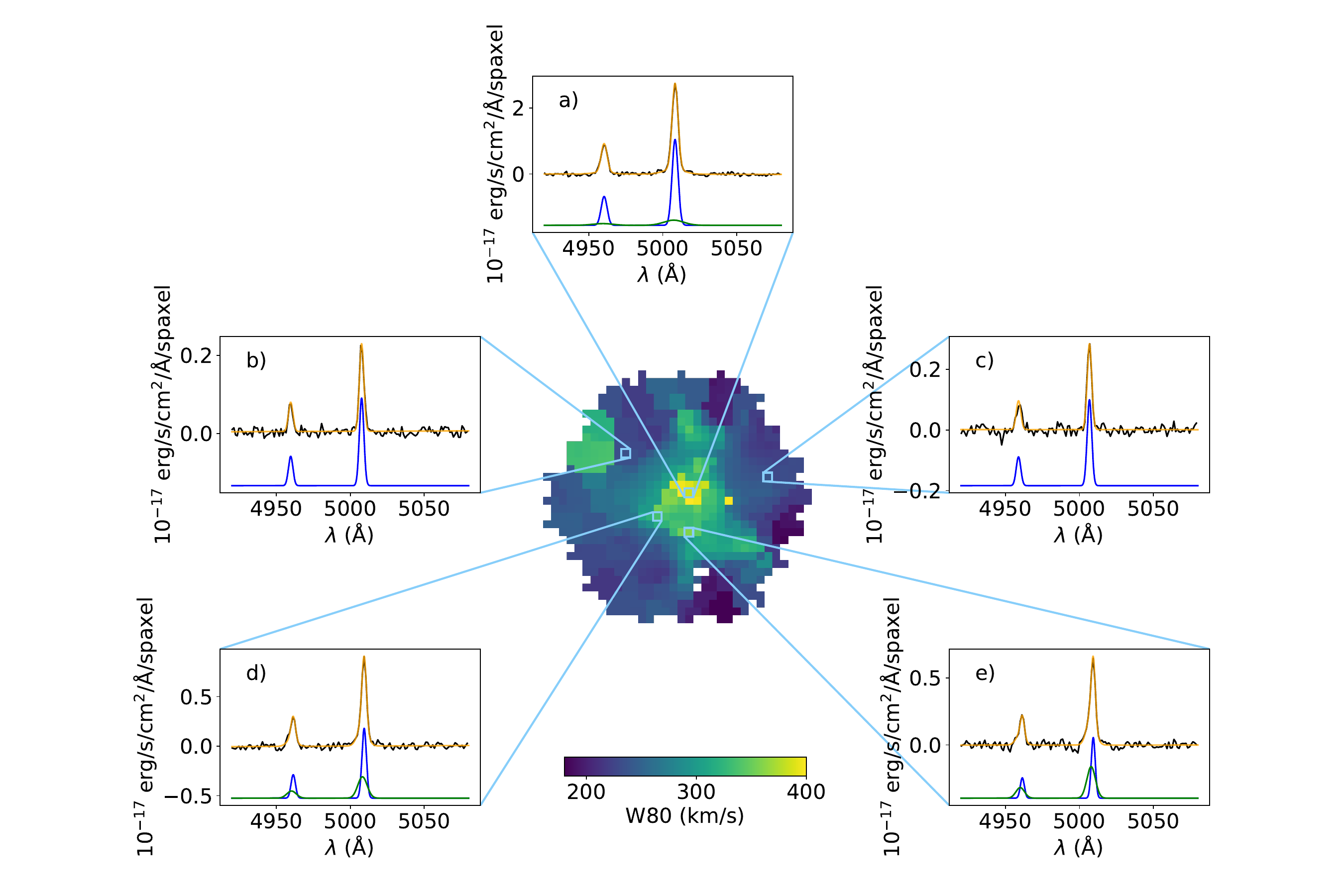}
    \caption{Examples of our fitting procedure in individual spaxels (for MaNGA plate-IFU: 9090-3704). The central image is the $W_{80}$ map for this target. The black line shows the original spectra, and the yellow line is our final fit. The blue and green lines are the Gaussian sub-components of the fit (they have been offset on the y-axis). If a panel has only one (blue) component (e.g., panel b or c), the single-Gaussian fit is best to model the spectrum.}
        \label{fig:representative_spectra}
\end{figure*}

In Figure \ref{fig:representative_spectra} we show an example of our fitting procedure in five spectra of one galaxy. We have particularly chosen a GHz-SF with high outflow rates (i.e., spaxels where our criterion chose two components to fit the [\ion{O}{iii}] profile). This example is representative of the behavior shown in Figure \ref{fig:out_frac}.

\begin{figure*}[h]
	\includegraphics[width=522pt]{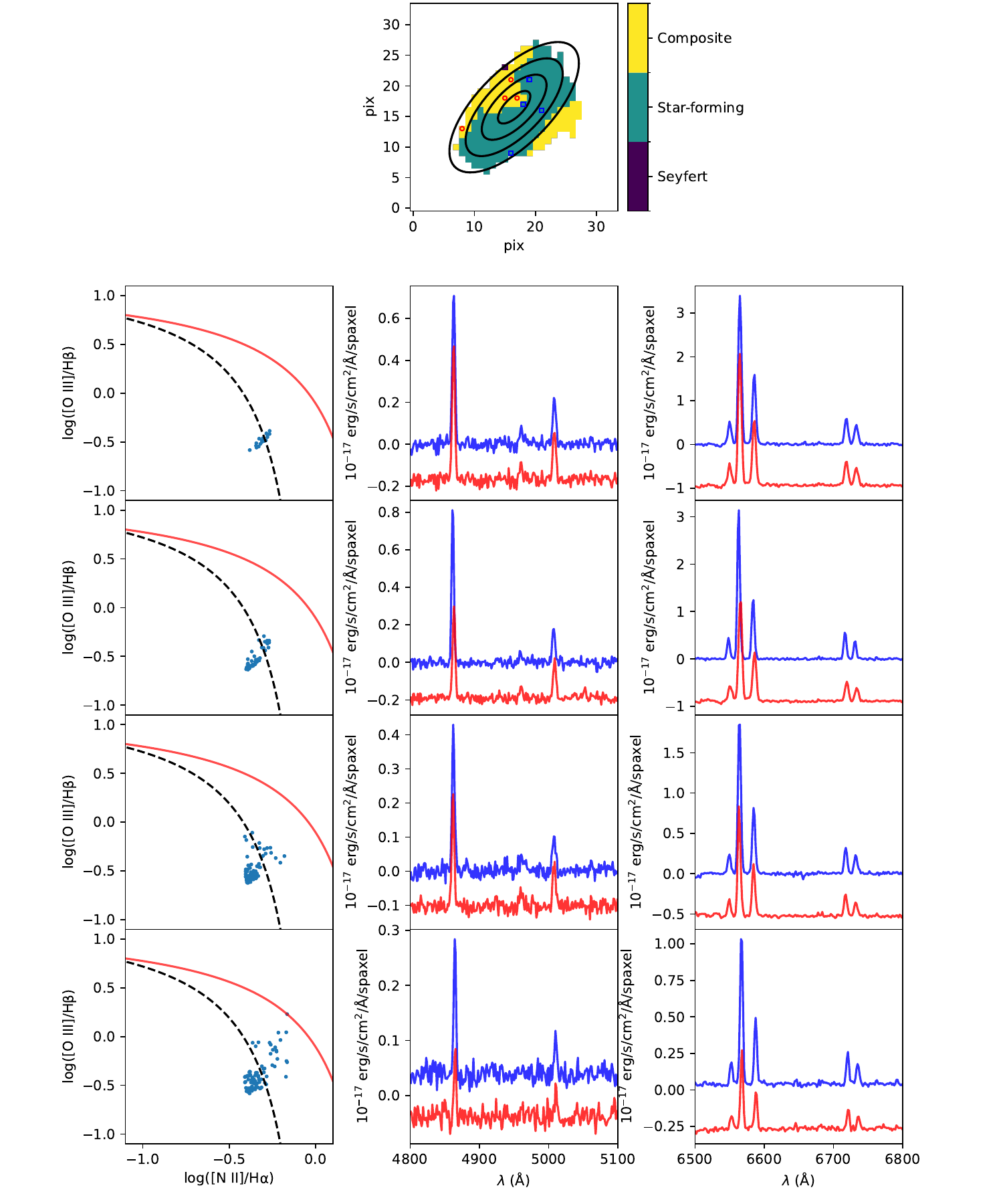}
    \caption{Spectra from individual spaxels at different annuli (for MaNGA plate-IFU: 8950-1901). The top plot illustrates the optical classification map based on the [\ion{N}{ii}]~BPT diagram following \citet{kewley_2006}. The black ellipses display annuli spaced at 0.5~$R_{eff}$. The four subsequent rows represent, at different annuli (as in Figure \ref{fig:bpt_contour}), the distribution of the [\ion{N}{ii}]~BPT diagram and spectra zoomed into two spectral regions containing the lines responsible for the classification. The exact spaxel where the spectra is taken from at each annuli is marked in the top panel (using a square for the blue spectra, and a circle for the red ones). The red spectra is offset in the y-axis.}
        \label{fig:representative_spectra_figBPT}
\end{figure*}

Similarly, in Figure \ref{fig:representative_spectra_figBPT} we show examples of different spectra from specific spaxels of a galaxy that contributes to Figure \ref{fig:bpt_contour}. For a single GHz-SF galaxy, on the top panel we show the resolved optical classification (into composite, star-forming or Seyfert) of each spaxel according to the [\ion{N}{ii}] BPT diagram \citep[following the guidelines from][]{kewley_2006}. Following the latter plot, the four rows of plots show representative data at four different annuli, as in Figure \ref{fig:bpt_contour}. Each row has three columns, which correspond to: the scatter through the [\ion{N}{ii}]~BPT diagram, a spectrum zoomed into the [\ion{O}{iii}] and H$\beta$ spectral region, and a spectrum zoomed into the [\ion{N}{ii}] and H$\alpha$ spectral region. In each row, the blue spectra shown in the two left panels correspond to a single, randomly selected spaxel. The same procedure is applied for the red spectrum (which is negatively offset in the y-axis). The spectra shown in Figure \ref{fig:representative_spectra} and Figure \ref{fig:representative_spectra_figBPT} are all stellar subtracted, taken from the MaNGA-DAP.

\end{appendix}

\end{document}